\documentclass[12pt]{article}
\usepackage{amsfonts}
\usepackage{latexsym}
\usepackage{dsfont}
\usepackage{amsmath}
\usepackage{amssymb}
\usepackage{amssymb}
\usepackage{amsthm}
\usepackage[nosort]{cite}
\usepackage{setspace}
\usepackage{graphicx}
\usepackage[font=small,labelfont=bf]{caption}
\usepackage[utf8]{inputenc}
\usepackage{geometry}
\usepackage[toc,page]{appendix}

\def\dd{\text{d}}

\def\la{\lambda}
\def\lb{\overline{\lambda}}
\def\ta{\tau}
\def\tb{\overline{\tau}}
\def\eb{\overline{e}}

\DeclareMathOperator\arccoth{arccoth}

\usepackage[linktocpage=true]{hyperref}
\hypersetup{
colorlinks=false,
citecolor= blue,
linkcolor= blue,
urlcolor= blue,
breaklinks=true
}

\def\appendix#1{\addtocounter{section}{1}\setcounter{equation}{0}
\renewcommand{\thesection}{\Alph{section}}
\section*{Appendix \thesection\protect\indent \parbox[t]{11.15cm}{#1}}
\addcontentsline{toc}{section}{Appendix \thesection\ \ \ #1}}

\allowdisplaybreaks

\begin{document}


\begin{titlepage}
\begin{center}
\vspace*{-1.0cm}
\hfill {\footnotesize HU-EP-21/35,  DMUS-MP-21/15}

\vspace{2.0cm}

{\LARGE  {\fontfamily{lmodern}\selectfont \bf Classical string solutions in
Non-Relativistic  \\
\vspace{3mm} AdS$_5\times$S$^5$: Closed and Twisted sectors}} \\[.2cm]

\vskip 2cm
\textsc{Andrea Fontanella\footnote{\href{mailto:andrea.fontanella@physik.hu-berlin.de}{\texttt{andrea.fontanella@physik.hu-berlin.de}}}$^{\Re}$ \footnotesize
and \normalsize Juan Miguel Nieto Garc\'ia\footnote{\href{mailto:j.nietogarcia@surrey.ac.uk}{\texttt{j.nietogarcia@surrey.ac.uk}}}$^{\Im}$\footnotesize }\\
\vskip 1.2cm

\begin{small}
\textit{$^{\Re}$Institut f\"ur Physik, Humboldt-Universit\"at zu Berlin, \\
IRIS Geb\"aude, Zum Gro{\ss}en Windkanal 2, 12489 Berlin, Germany}

\vspace{5mm}

\textit{$^{\Im}$Department of Mathematics,
University of Surrey \\
Guildford, GU2 7XH, UK. }
\end{small}

\end{center}

\vskip 1 cm
\begin{abstract}
\vskip1cm
 
We find classical closed string solutions to the non-relativistic AdS$_5\times$S$^5$ string theory which are the analogue of the BMN and GKP solutions for the relativistic theory. 
We show that non-relativistic AdS$_5\times$S$^5$ string theory admits a $\mathbb{Z}_2$ orbifold symmetry which allows us to impose twisted boundary conditions. Among the solutions in the twisted sector, we find the one around which the semiclassical expansion in \href{https://arxiv.org/abs/2102.00008}{arXiv:2102.00008} takes place.   

\end{abstract}

\end{titlepage}

\tableofcontents
\vspace{5mm}
\hrule


 \setcounter{section}{0}
 \setcounter{footnote}{0}

\section*{Introduction}\addcontentsline{toc}{section}{Introduction}

In Modern Physics, laws of Nature are typically written in a relativistic formalism, where spacetime is modelled with a Lorentzian manifold.  However, many physical systems can be effectively described by a non-relativistic (NR) action. The most common examples are multi-body problems in condensed matter, but they are not the only ones.  Non-relativistic Physics also appears in gravitational phenomena, such as black holes scattering, where General Relativity (GR) is well approximated by the Post-Newtonian expansion (PN).  

In the recent years, there has been an intense study to find the proper mathematical description of non-relativistic General Relativity (NRGR) \cite{Festuccia:2016awg, Hansen:2018ofj, Hansen:2020pqs, Hansen:2019vqf, Bergshoeff:2017dqq, Bergshoeff:2017btm}.  In these studies, it has been found that \emph{Newton-Cartan} (NC) geometry has the right properties to model spacetime in a covariant manner in the context of NRGR.   NC geometry was formulated by E. Cartan in 1923 \cite{Cartan1, Cartan2}, following the success of Einstein's discovery of GR,  as a way to geometrically describe Newton's law.  For long time it has been forgotten (see \cite{Trautman, Havas} for some works in the 60s), but recently there has been a revival, mostly because of modern advances in the field of high-energy physics that gave the motivation to study this topic. 

The PN expansion of GR turns out particularly helpful to compute the scattering amplitude for collision of the binary system of black holes \cite{Levi:2018nxp}.  
The PN expansion is a low-velocity and weak-field expansion of GR,  where $v/c$ and the Newton's constant $G_N$ are taken to be very small. 
Understanding the mathematical structure of NRGR, based on NC geometry, would provide the necessary theoretical ground for these types of computations, where order by order in $v/c$, the perturbative effects in $G_N$ are automatically resummed into the theory.

Because NRGR appears as a limit of the well-establish theory of GR, it naturally gives us a route to look into quantum gravity from a different perspective.   
In particular, it is an important question to ask whether NRGR admits a UV completion. This can be answered, for instance,  using the framework of non-relativistic string theory (NRST). 

NRST was formulated for the first time by Gomis and Ooguri in flat space  \cite{Gomis:2000bd}, and later in AdS$_5\times$S$^5$ by Gomis, Gomis and Kamimura \cite{Gomis:2005pg}.  The NRST proposed by these authors is a 2d string sigma model, where the world-sheet is relativistic, and the spacetime is NR.  It was found later that the spacetime is described by \emph{string} Newton-Cartan (sNC) geometry \cite{Harmark:2017rpg, Harmark:2018cdl, Harmark:2019upf, Bergshoeff:2018yvt, Bergshoeff:2019pij, Gomis:2019zyu}, and that NRST can be generalised to any type of curved backgrounds. There exists two different but complementary approaches to NRST: one consists of a large-$c$ expansion of the vielbein for the spacetime metric,  and the other one on the null-reduction of a Riemannian manifold.  As discussed in \cite{Harmark:2019upf}, the two approaches are equivalent.

There is also the possibility of taking a double scaling limit which makes both the target space and the world-sheet non-relativistic. This setting has been shown to be the natural framework to describe e.g. the Spin Matrix limit of AdS$_5\times$S$^5$ \cite{Harmark:2017rpg, Harmark:2018cdl, Harmark:2020vll}.  
Classical spinning string solutions for this type of theory have been studied in e.g. \cite{Roychowdhury:2019olt, Roychowdhury:2019sfo, Roychowdhury:2020kma, Roychowdhury:2020dke,   Roychowdhury:2020yun, Roychowdhury:2021wte}.

In this paper, however, we shall not discuss non-relativistic strings with non-relativistic world-sheet,  and every time  we talk about NRST we refer to theory akin to the one formulated by Gomis and Ooguri.         

So far, only the bosonic sector of NRST has been properly understood in the formalism of sNC geometry.  It is still an open question how to couple fermions to this type of sigma model and whether there is a low energy effective action, namely a NR supergravity,  which governs the dynamics of the background fields. For recent work in this direction, see \cite{NRsugra}.

Another important motivation to study NRST is due to the holographic principle.  
Currently, the holographic principle is mainly understood via the AdS/CFT correspondence\cite{Aharony:1999ti}, however very little is known about non-AdS holography. The fact that NRST is based on a non-Riemannian geometry while retaining part of the conformal symmetry of AdS backgrounds,  makes it an appealing arena where a NR holographic principle of non-AdS type could be formulated.  It is still an open question whether in this context there is a notion of near-horizon geometry which decouples from the bulk and how to identify the boundary of a sNC manifold.  For past works related to NR holography, see e.g. \cite{Taylor:2008tg, Taylor:2015glc, Costa:2010cn, Guica:2010sw, Hartong:2013cba, Christensen:2013lma, Christensen:2013rfa, Hartong:2014oma, Hartong:2014pma, Dobrev:2013kha, Sakaguchi:2007ba}

An interesting property of relativistic AdS$_5\times$S$^5$ is that its equations of motion admit a Lax pair formulation,  leading to classical integrability of the theory \cite{Bena:2003wd}.  The fact that the theory is exactly solvable (not only at the classical level, but also at the quantum level) leads to important results confirming the holographic conjecture by matching the spectrum of string excitations with conformal dimensions of corresponding gauge invariant operators in the dual CFT, see \cite{Beisert:2010jr} for a review. 

At the moment of writing, it is still an open question whether NRST admits a Lax pair formulation and work is in progress in this direction \cite{Andrea_Stijn}.  In \cite{Fontanella:2020eje}, the method of Lie algebra expansion has been used to generate new integrable sigma models, and an integrable extension of the theory \cite{Gomis:2005pg}, with supersymmetry, has been obtained.  Whether this extended theory can be consistently truncated to the one constructed in \cite{Gomis:2005pg} and if such truncation would still retain the Lax pair formulation are open questions.

A Lax pair for the theory obtained by rescaling the AdS$_5$ time and the angle of the greater circle in S$^5$ was found in \cite{Kluson:2017ufb}. However this limit is not the same NR limit of \cite{Gomis:2005pg}, since it is performed on the two isometry directions of the background, and it does not single out an AdS$_2$ subspace inside AdS$_5$.    
Another claiming of constructing a Lax pair for the theory in \cite{Gomis:2005pg} was given in \cite{Roychowdhury:2019vzh}. However in this paper there is the assumption that the Maurer-Cartan 1-form takes values in the non-central extension of string Galilei algebra, known as Bargmann algebra \cite{Bergshoeff:2018yvt}. From the coset space construction, this is secretly assuming that the target space is the NR flat spacetime. Therefore the Lax pair so constructed may prove classical integrability for NR strings in flat spacetime (which in static gauge reduces to free fields in 2d Minkowski, therefore trivially integrable), but not for the non-trivial dynamics occurring for NR strings in AdS$_5\times$S$^5$.

Another important object which one would like to compute is the S-matrix.  If classical integrability was known, then one could use symmetries to construct the exact S-matrix.  In general,  however, one needs a well-defined perturbative expansion of the theory, which in turn one can use to compute the S-matrix perturbatively. An issue one encounters for the NRST compared with the relativistic case is that the action heavily simplifies if we choose an AdS$_2$ world-sheet instead of a flat wordsheet. However, asymptotic states are not well defined for curved spaces.\footnote{As far as we are aware of, the only way to define something similar to asymptotic states for AdS$_2$ requires to enlarge the space with additional patches \cite{Anninos:2019oka}.}
A semiclassical expansion in large string tension of NRST in AdS$_5\times$S$^5$ has been found in \cite{Fontanella:2021hcb} for a flat wordsheet.  The procedure followed in that work consists in fixing the light-cone gauge\footnote{For other works related to gauge fixing and Hamiltonian formulation of NRST, see e.g. \cite{Kluson:2017ufb, Kluson:2018grx, Kluson:2018egd, Kluson:2017abm}.} and expand the action for  large string tension and also for large common AdS$_5$ and S$^5$ radius.
The fields are rescaled around the null value, except for the NR longitudinal direction, which needs to be shifted by a linear function in the string coordinate $\sigma$. Although in this case asymptotic states are, in principle, well defined, we find a non-trivial $\sigma$-dependence in the sub-leading Lagrangian corrections. Although this dependence is controlled by the radius of the geometry and can be ignored in the large-radius regime, it creates an obstruction to the computation of the $S$-matrix. This happens because the large-radius limit is incompatible with the world-sheet decompactification limit, where the $S$-matrix is well defined. In addition, a complete understanding of the classical string solution around which this semiclassical expansion happens was still missing. It is part of the motivation of this paper to elucidate this point and to show explicitly the classical solution around which the expansion in \cite{Fontanella:2021hcb} takes place, namely the twisted BMN-like solution presented in section \ref{twisted_BMN_sol}.

Classical string solutions have been widely studied in relativistic string theory, among other reasons, due to their relevance in the context of the AdS/CFT correspondence. The dispersion relation of semiclassical states described by solitonic closed-string solutions can be matched with the dimensions of the corresponding gauge-theory operator in the limit of large quantum numbers, even beyond BPS operators \cite{Beisert:2003xu,Frolov:2003xy}. The accuracy of the comparison is controlled by the sigma model coupling constant, as quantum corrections are suppressed when it is large. Prime examples of classical solutions are the BMN string \cite{Berenstein:2002jq} and the GKP string \cite{Gubser:2002tv,Frolov:2002av}. The BMN is a point-like string (particle) moving with the speed of light along a great circle of the S$^5$ space, while the GKP string is a rigid string moving in AdS$_3\subset$ AdS$_5$. Multi-spin string configurations were later studied in \cite{Frolov:2002av,Frolov:2003qc}. There exist many more interesting solutions, but they will not be relevant for this particular article. We refer the reader to \cite{0311139,Tseytlin:2010jv} for a review on the subject. We will draw inspirations from these results when writing ans\"atze for classical string solutions of NRST in AdS$_5\times$S$^5$ of \cite{Gomis:2005pg}.

\vspace{2mm}
\emph{Plan of this paper.} In section \ref{sec:rel_sol} we review some concepts of classical string solutions in relativistic AdS$_5\times$S$^5$ string theory. In section \ref{sec:non_rel_general} we introduce the notion of non-relativistic string theory based on a general Newton-Cartan geometry, with some emphasis on the AdS$_5\times$S$^5$ case. A $\mathbb{Z}_2$ orbifold symmetry of the theory will also be shown.
In section \ref{sec:clas_sol} we study three different types of classical solutions appearing in NR AdS$_5\times$S$^5$ string theory. Closed and twisted boundary conditions will be imposed, leading to two different sectors of solutions. Two of those closed string solutions can be understood as non-relativistic analogues of the celebrated BMN and GKP solutions. 
We end this article with a summary, conclusions and future prospects.

\section{Classical solutions in relativistic AdS$_5\times$S$^5$ string theory}
\label{sec:rel_sol}
\setcounter{equation}{0}

\subsection{Relativistic string theory action}

Our starting point is to consider the bosonic part of the classical string action in AdS$_5\times$S$^5$ space-time
\begin{equation}
\label{rel_action}
S = -\frac{T}{2} \int \dd^2 \sigma \,  \bigg(\gamma^{\alpha\beta} \partial_{\alpha} X^{\mu} \partial_{\beta} X^{\nu} g_{\mu\nu} + \varepsilon^{\alpha\beta}  \partial_{\alpha} X^{\mu} \partial_{\beta} X^{\nu} b_{\mu\nu} \bigg) \ , 
\end{equation}
where $T$ is the string tension, $\sigma^{\alpha} = (\tau, \sigma)$, with $\alpha = 0, 1$, are the string world-sheet coordinates, $\gamma^{\alpha\beta} \equiv \sqrt{-h} h^{\alpha\beta}$ is the Weyl invariant combination of the inverse world-sheet metric $h^{\alpha\beta}$ and $h =$ det$(h_{\alpha\beta})$. $b_{\mu\nu}$ is a \emph{closed} Kalb-Ramond field ($\dd b = 0$), which we set to zero for the moment, but it will be relevant later in the context of the non-relativistic limit.  $g_{\mu\nu}$ is the AdS$_5\times$S$^5$ metric.  $X^{\mu}$ are the string embedding coordinates and they should be understood as bosonic fields depending on $(\tau, \sigma)$.

Among the different coordinate systems we can use, both the real and complex embedding coordinates prove to be very convenient to describe classical strings. First, we define the real 6-dimensional embedding coordinates of the sphere $S^5$ as the ones describing the following surface in $\mathbb{R}^6$
\begin{equation}
	x_M x_M=x_1^2 + \dots +x_6^2 =R^2 \ ,
\end{equation}
while the $AdS_5$ space coordinates can be represented similarly in a $\mathbb{R}^{2,4}$ as
\begin{equation}
	-\eta^{MN} y_M y_N=y_0^2 -y_1^2 + \dots -y_4^2 +y_5^2 =R^2 \ .
\end{equation}
Notice that the radius $R$ is common for the AdS$_5$ and S$^5$ spaces. The complexified coordinates are obtained from those as $\mathbb{X}_j=x_{2j-1}+i x_{2j}$, $\mathbb{Y}_0=y_0 + i y_5$ and $\mathbb{Y}_j=y_{2j-1}+i y_{2j}$. In both sets of coordinates the $SO(2,4)\times SO(6)$ symmetry of the space is explicit, 
 and we can define the following conserved quantities
\begin{equation}
	S_{PQ}=-T \int_0^{2\pi} \dd\sigma (y_P \dot{y}_Q- y_Q \dot{y}_P ) \ , \qquad J_{MN}=-T \int_0^{2\pi} \dd\sigma (x_M \dot{x}_N- x_N \dot{x}_M ) \ .
\end{equation}
There is a natural choice of the 3+3 Cartan elements of this symmetry as the generators associated to translations of the phases of the complexified coordinates
\begin{equation}
	E=S_{50} \ , \qquad S_1=S_{12} \ , \qquad S_2=S_{34} \ , \qquad J_1=J_{12} \ , \qquad J_2=J_{34} \ , \qquad J_3=J_{56} \ .
\end{equation}
These are usually called \emph{energy}, \emph{spin} and \emph{angular momenta} respectively.

Because the target space we are interested in is the tensor product of two spaces, working in conformal gauge proves to be incredibly convenient. The reason is that, by choosing $h^{\alpha \beta}=\eta^{\alpha \beta}$, the degrees of freedom of the two subspaces do not intertwine in the action and it can be split into an action for the AdS degrees of freedom and an action for the sphere degrees of freedom
\begin{gather}
	\mathcal{L}= \mathcal{L}_{AdS} + \mathcal{L}_S \ , \label{rel_lag}\\
	\mathcal{L}_{AdS} = - \partial^a y^P \partial_a y_P -\hat{\Lambda} (y_P y^P +1) \ , \qquad \mathcal{L}_{AdS} = - \partial^a x_M \partial_a x_M -\Lambda (x_M x_M -1) \ . \notag
\end{gather}
However, this does not mean that the two sets of degrees of freedom completely decouple. The string theory action in the conformal gauge has to be supplemented with the Virasoro constraints, which force the two-dimensional stress-energy tensor to vanish
\begin{equation}
\dot{y}^P \dot{y}_P + y^{\prime P} y'_P+\dot{x}_M \dot{x}_M + x'_M x'_M=0 \ , \qquad \dot{y}^P y'_P +\dot{x}_M x'_M=0 \ .
\end{equation}

Finally, we should comment that the conformal gauge does not eliminate all the redundancy of the action. In fact, it reduces the original 2-dimensional diffeomorphisms in $\tau$ and $\sigma$, which we will denote by  Diff$_2$, to separate diffeomorphisms groups action on the light-cone coordinates $\sigma_\pm=\tau\pm \sigma$, which we will denote by Diff$_+ \oplus$Diff$_-$. This remnant of redundancy is usually eliminated by demanding the phase of $\mathbb{Y}_0$ to be equal to $\kappa \tau$.

\subsection{Classical rigid strings}

The equations of motion of the Lagrangian (\ref{rel_lag}) are
\begin{align}
	\partial^a \partial_a y_Q -\hat{\Lambda} y_Q &=0 \ , & y_Q y^Q &=-1 \ ,  \\
	\partial^a \partial_a x_M +\Lambda x_M &=0 \ , & x_M x_M &=1 \ .
\end{align}

The two simplest solutions we can write are point-like strings, i.e., strings that are independent of the $\sigma$ coordinate. Up to global $SO(2,4)\times SO(6)$ transformations, there exists two solutions of this kind: the massless AdS geodesic, which is a straight line in $\mathbb{R}^{2,4}$, and the BMN string, which runs along the time direction of the $AdS_5$ and wraps a big circle of $S^5$. No other solution of this kind exists because the Virasoro constraints force $\hat{\Lambda}=-\Lambda$ once the equations of motion are taken into account. Here we will be interested only in the latter solution.\footnote{Although the massless AdS geodesic is also an interesting solution, it does not represent a ``good semiclassical string state'' because it always has non-vanishing non-Cartan components of the angular momentum \cite{0311139}.}

The BMN string can be written as $\mathbb{Y}_0=\mathbb{X}_3=e^{i\kappa \tau}$, with the remaining coordinates set to zero. The phase of $\mathbb{Y}_0$ can be mapped to the time coordinate of the AdS$_5$ space, while the phase of $\mathbb{X}_3$ parametrizes a big circle on S$^5$. Thus, in some appropriate coordinates, this string can be described as $t=\phi=\kappa \tau$. The non-trivial conserved charges associated to this string configuration are $E=J_{3}=-2\pi T\sqrt{\Lambda}=-2\pi T\kappa$. This classical solution is related via holography to the protected operator tr$(Z^{J_3})$ of the $\mathcal{N}=4$ SYM field theory.

The BMN solution can be generalized into a multi-spin solutions by allowing non-vanishing values of the other two Cartan angular momenta. The Lagrangian associated to these configurations can be shown to be related to the 1-dimensional Neumann-Rosochatius mechanical system \cite{Arutyunov:2003uj,Arutyunov:2003za}. The Neumann-Rosochatius model describes an oscillator on a sphere with a centrifugal potential, and it is known to be integrable.

Apart from rotating strings in S$^5$, another important example of a non-trivial rigid string solution is the one that describes a folded spinning string in AdS$_3$. Prime example of those is the GKP string. The GKP string can be written as $\mathbb{Y}_0=\cosh \rho \, e^{i\kappa \tau}$, $\mathbb{Y}_1= \sinh \rho \, e^{i\omega \tau}$, with the remaining coordinates set to zero. The $\rho$ coordinate we have introduced is a real function of $\sigma$ fixed by the equations of motion. In addition, it is bounded between $0$ and $\rho_0=\arccoth \frac{\omega}{\kappa}$. There exist two interesting regimes of this solution: the short string, associated to $\rho_0\rightarrow 0$, and the long string, associated to $\rho_0 \rightarrow \infty$. In the former case we find that the energy goes as $E\propto \sqrt{S}$, while the latter case is more subtle, giving us $E=S+2\,T \ln (\tfrac{S}{2\pi T}) + \dots$.

\section{Non-relativistic AdS$_5\times$S$^5$ string theory}
\label{sec:non_rel_general}
\setcounter{equation}{0}

\subsection{Non-relativistic string theory action}
In order to construct the non-relativistic AdS$_5\times$S$^5$ string action, we shall begin with the relativistic one (\ref{rel_action}) and take a limit on the rescaled coordinates dictated by the contraction of the AdS$_5$ isometry algebra to the 5d Newton-Hooke.  We shall specifically choose Cartesian global coordinates to work with\footnote{The construction of the non-relativistic limit is heavily dependent on the coordinate parameterisation we use for describing our space. This point is studied with some level of detail in the Appendix A of \cite{Fontanella:2021hcb}.} i.e.  $X^{\mu} = (t, z_i, \phi, y_i)$, where $i=1,..., 4$. For these coordinates, the metric takes the form
\begin{equation}
\label{metric_cartesian}
g_{\mu\nu} \dd X^{\mu} \dd X^{\nu} = - \bigg(\frac{1+ \frac{z^2}{4 R}}{1-\frac{z^2}{4 R}}\bigg)^2 \dd t^2 + \frac{1}{(1-\frac{z^2}{4 R})^2} \dd z_i \dd z_i + \bigg(\frac{1 - \frac{y^2}{4 R}}{1 + \frac{y^2}{4 R}}\bigg)^2\dd \phi^2 + \frac{1}{(1+\frac{y^2}{4 R})^2} \dd y_i \dd y_i \ ,
\end{equation}
where again $R$ is the common radius of AdS$_5$ and S$^5$. We use the short-hand notation $z^2\equiv z_i z^i$, where $z^i = z_i$, and the same for the $y$ coordinates.  
The range of coordinates is $-\infty < z_i, y_i < + \infty$ and the AdS time is periodic $0< t \leq 2\pi$. However, in order to avoid closed time-like curves, we shall take the universal cover of AdS, where $-\infty < t < + \infty$. The coordinate $\phi$ spans the circle in S$^5$, $0< \phi \leq 2\pi R$.  
In the limit $R \rightarrow \infty$, the action (\ref{rel_action}) with metric (\ref{metric_cartesian}) describes strings moving in flat spacetime. 

Following the procedure suggested in \cite{Fontanella:2021hcb},  the contraction of the AdS$_5$ isometry algebra $\mathfrak{so}(4,2)$ to the 5d Netwon-Hooke algebra imposes the following rescaling of the coordinates
\begin{equation}
\label{coord_resc}
t \rightarrow t \ ,  \qquad 
z_1 \rightarrow z_1 \ ,  \qquad
 z_m \rightarrow \frac{1}{c} z_m \ , \qquad
 \phi \rightarrow \frac{1}{c} \phi \ ,  \qquad
 y_i \rightarrow \frac{1}{c} y_i \ , 
\end{equation}
where $m=2,3,4$. In addition, we rescale the string tension as
\begin{equation}
\label{tension_resc}
T \rightarrow c^2\,  T \ . 
\end{equation}
Here $c$ is a non-relativistic contraction parameter, which plays the r\^ole of the speed of light and is assumed to be large. 
We remark that the NR limit (\ref{coord_resc}) already takes into account the contribution from the radius rescaling $R \rightarrow c R$ originally introduced in \cite{Gomis:2005pg}. Instead, as described in Appendix A of \cite{Fontanella:2021hcb}, the radius inside the metric (\ref{metric_cartesian}) does not need to be rescaled with $c$, and the limit (\ref{coord_resc}) is exactly the limit taken in \cite{Gomis:2005pg}. However, the radius defining the periodicity of $\phi$ needs to be rescaled by $c$, and a consequence of this\footnote{We thank J. Kluso\v{n} for mentioning this point.}, is that it becomes infinite when $c \rightarrow \infty$, i.e. $\phi$ spans $\mathbb{R}$.

If we write the spacetime metric, taking into account the additional $c^2$ factor from the string tension rescaling, as 
\begin{equation}
G_{\mu\nu} \equiv c^2 g_{\mu\nu} = \hat{E}_{\mu}{}^{\hat{A}} \hat{E}_{\nu}{}^{\hat{B}} \hat{\eta}_{\hat{A}\hat{B}} \ , 
\end{equation}
then the coordinates rescaling (\ref{coord_resc}) implies that the vielbeine expands in large $c$ as follows
\begin{eqnarray}
\label{vielbeine_exp}
 \hat{E}_{\mu}{}^A = c \tau_{\mu}{}^A + \frac{1}{c} m_{\mu}{}^A + \mathcal{O}(c^{-2})\ , \qquad\qquad
 \hat{E}_{\mu}{}^a = e_{\mu}{}^a + \mathcal{O}(c^{-1}) \ , 
\end{eqnarray}
where $\hat{A} = 0, ... , d-1$ is decomposed into $A= 0, 1$ (longitudinal) and  $a = 2, ..., d-1$ (transverse). We will refer to the set $\{\tau_{\mu}{}^A, m_{\mu}{}^A, e_{\mu}{}^a\}$ as Newton-Cartan data. In our case, it turns out to be
\begin{eqnarray}
\notag
\tau_{\mu}{}^A &=& \text{diag}\bigg(-\frac{1+ (\frac{z_1}{2R})^2}{1- (\frac{z_1}{2R})^2},   \frac{1}{1 - (\frac{z_1}{2R})^2}, 0, ... , 0 \bigg) \ , \\
m_{\mu}{}^A &=& \text{diag}\bigg(-\frac{z_m z_m}{2R^2(1-(\frac{z_1}{2R})^2)},  \frac{z_m z_m}{4R^2(1-(\frac{z_1}{2R})^2)}, 0, ... , 0 \bigg) \ , \\
\notag
e_{\mu}{}^a &=&\text{diag} \left(0, 0, \frac{1}{1 - (\frac{z_1}{2R})^2}, \frac{1}{1 - (\frac{z_1}{2R})^2}, \frac{1}{1 - (\frac{z_1}{2R})^2}, 1,1,1,1,1 \right) \ .
\end{eqnarray}
The transverse vielbeine $e_{\mu}{}^a$ extracted from the Maurer-Cartan 1-form in these coordinates do not appear in this diagonal form \cite{Fontanella:2021hcb}. Here we performed an additional local rotation in the transverse space to write them in this form.
The coordinates $t$ and $z_1$ are longitudinal time-like and space-like directions, respectively\footnote{In the sense that their associated vector fields $k = \partial_t$ and $\ell = \partial_{z_1}$ satisfy the conditions 
\begin{equation*}
\tau_{\mu}{}^0 k^{\mu} \neq 0 \ , \qquad
\tau_{\mu}{}^1 k^{\mu} = 0 \ , \qquad
e_{\mu}{}^a k^{\mu} = 0 \ , 
\end{equation*}
and
\begin{equation*}
\tau_{\mu}{}^0 \ell^{\mu} = 0 \ , \qquad
\tau_{\mu}{}^1 \ell^{\mu} \neq 0 \ , \qquad
e_{\mu}{}^a \ell^{\mu} = 0 \ .
\end{equation*}}.

The vielbeine expansion, in turn, induces the expansion of the action (\ref{rel_action}) as
\begin{equation}
\label{S_NR_div}
S = - \frac{T}{2} \int \dd^2 \sigma \, \gamma^{\alpha\beta} \bigg( c^2 \partial_{\alpha} X^{\mu} \partial_{\beta} X^{\nu} \tau_{\mu\nu} + \partial_{\alpha} X^{\mu} \partial_{\beta} X^{\nu} H_{\mu\nu} + \varepsilon^{\alpha\beta}  \partial_{\alpha} X^{\mu} \partial_{\beta} X^{\nu} B_{\mu\nu}\bigg) + \mathcal{O}(c^{-2}) \ , 
\end{equation} 
where
\begin{equation}
\tau_{\mu\nu} \equiv \tau_{\mu}{}^A \tau_{\nu}{}^B \tilde{\eta}_{AB} \ , \qquad\qquad
H_{\mu\nu} \equiv e_{\mu}{}^a e_{\nu}{}^b \tilde{\delta}_{ab} + \bigg(\tau_{\mu}{}^A m_{\nu}{}^B + \tau_{\nu}{}^A m_{\mu}{}^B \bigg) \tilde{\eta}_{AB} \ , 
\end{equation}
with $\tilde{\eta}_{AB} =$ diag$(-1, 1, 0, ..., 0)$, $\tilde{\delta}_{ab} =$ diag$(0,0,1,...,1)$.  
In our specific case,  $\tau_{\mu\nu}$ (\emph{longitudinal} metric) is
\begin{equation}
\tau_{\mu\nu} \dd X^{\mu} \dd X^{\nu} = - \bigg(\frac{1+ (\frac{z_1}{2R})^2}{1- (\frac{z_1}{2R})^2}\bigg)^2 \dd t^2 + \frac{1}{(1 - (\frac{z_1}{2R})^2)^2} \dd z_1^2 \ ,
\end{equation}
which describes an AdS$_2$ subspace,  and $H_{\mu\nu}$ (\emph{boost invariant} metric)  is 
\begin{eqnarray}
\notag
H_{\mu\nu} \dd X^{\mu} \dd X^{\nu} &=& - \frac{(1+ (\frac{z_1}{2R})^2)z_m z_m}{R^2(1- (\frac{z_1}{2R})^2)^3} \, \dd t^2 
+ \frac{z_m z_m}{2R^2(1- (\frac{z_1}{2R})^2)^3}\,  \dd z_1^2 \\
&+& \frac{1}{(1- (\frac{z_1}{2R})^2)^2} \, \dd z_m \dd z_m + \dd \phi^2
+ \dd y_i \dd y_i \ .
\end{eqnarray}
One should note that the two vector fields $\partial_t$ and $\partial_{\phi}$ are still isometries for the two metrics $(\tau_{\mu\nu},H_{\mu\nu})$ in the usual sense of Lie derivatives.  Moreover,  the coordinates $\{\phi, y_i\}$ now parametrise a 5d Euclidean space through the metric $H_{\mu\nu}$. 

When taking the contraction limit $c \rightarrow \infty$, the term proportional to $c^2 \tau_{\mu\nu}$ diverges. This is cured by turning on a closed Kalb-Ramond B-field given as follows\footnote{The convention is fixed to be $\varepsilon^{01} = + 1$ for both $\varepsilon^{\alpha\beta}$ and $\varepsilon^{AB}$.}
 \begin{equation}
B_{\mu\nu} \equiv  c^2 b_{\mu\nu} = c^2 \tau_{\mu}{}^A \tau_{\nu}{}^B \varepsilon_{AB} \ , 
\end{equation}
where we have included the rescaling of the string tension (\ref{tension_resc}). 
In this way the divergent part of the action can be recast into a perfect square
\begin{equation}
\label{action_F2}
S  = - \frac{T}{2} \int \dd^2 \sigma \bigg( \, \gamma^{\alpha\beta}\partial_{\alpha} X^{\mu} \partial_{\beta} X^{\nu} H_{\mu\nu} + c^2 \, \gamma^{00} \mathcal{F}^A \mathcal{F}^B \tilde{\eta}_{AB} + \mathcal{O}(c^{-2}) \bigg)\ , 
\end{equation}
where
\begin{equation}
\label{F}
 \mathcal{F}^A = \tau_{\mu}{}^A \partial_0 X^{\mu} - \frac{1}{\gamma_{11}} \varepsilon^{AB} \tilde{\eta}_{BC} \tau_{\mu}{}^C \partial_1 X^{\mu} - \frac{\gamma_{01}}{\gamma_{11}} \tau_{\mu}{}^A \partial_1 X^{\mu} \ . 
\end{equation}
Finally, the perfect square divergent part can be rewritten in terms of a finite quantity in the $c\rightarrow \infty$ limit,  i.e. 
\begin{equation}
\label{rewriting}
- \frac{T c^2}{2} \int \dd^2 \sigma  \, \gamma^{00} \mathcal{F}^A \mathcal{F}^B \tilde{\eta}_{AB} = - \frac{T}{2} \int \dd^2 \sigma \bigg(  \lambda_A \mathcal{F}^A + \frac{1}{4 c^2 \gamma^{00}} \lambda_A \lambda^A\bigg) \ , 
\end{equation}
at the price of introducing Lagrange multipliers $\lambda_A = \lambda_A(\tau, \sigma)$, which should be regarded, before any gauge fixing, as world-sheet scalars.  To show the equivalence between the two sides of equation (\ref{rewriting}) one needs to insert back the equations of motion for the non-dynamical fields $\lambda_A$.
At this point we are allowed to take the limit $c \rightarrow \infty$, and by doing this we end up with the following action
\begin{equation}
\label{NR_action}
S^{NR} = - \frac{T}{2} \int \dd^2 \sigma \, \bigg( \gamma^{\alpha\beta}\partial_{\alpha} X^{\mu} \partial_{\beta} X^{\nu} H_{\mu\nu} + \lambda_A \mathcal{F}^A \bigg) \ . 
\end{equation}
This action describes strings propagating on a background which is non-Lorentzian, but a string Newton-Cartan manifold. The local symmetry is the string Newton-Cartan algebra, which is a particular (non-central) extension of the (string) Galilei algebra \cite{Bergshoeff:2019pij}\footnote{See also the recent work \cite{Bidussi:2021ujm, Hartong:2021ekg} for a new revisited formulation of the NR string action.}.  

The equations of motion for $\lambda_A$ imply that 
\begin{equation}
h_{\alpha\beta} = \varphi(\tau, \sigma) \, \tau_{\mu\nu} \partial_{\alpha} X^{\mu}   \partial_{\beta} X^{\nu} \equiv  \varphi(\tau, \sigma)\, \tau_{\alpha\beta} \ , 
\end{equation}
where $\varphi(\tau, \sigma)$ is a world-sheet scalar function.  By plugging this result back into the Polyakov action (\ref{NR_action}) one obtains the non-relativistic action in the Nambu-Goto form, 
\begin{equation}
\label{NR_action_NambuGoto}
S^{NR} =- \frac{T}{2} \int \dd^2 \sigma \sqrt{- \tau} \tau^{\alpha\beta} \partial_{\alpha} X^{\mu} \partial_{\beta} X^{\nu} H_{\mu\nu} \ . 
\end{equation}

\noindent\textbf{An alternative rewriting.} An alternative but equivalent way to write the Lagrange multiplier term appearing in the Polyakov non-relativistic action was proposed in \cite{Bergshoeff:2018yvt}. Here we remark that this rewriting is correct, but under a certain assumption regarding the world-sheet Zweibeine, which will be given later in equation (\ref{positivity_Zweibeine}).  We begin by defining 
\begin{equation}
\mathcal{F} \equiv \mathcal{F}^0 +   \mathcal{F}^1 \ , \qquad\qquad
\overline{\mathcal{F}} \equiv \mathcal{F}^0 - \mathcal{F}^1 \ , 
\end{equation}
such that 
\begin{equation}
\mathcal{F}^2 \equiv\mathcal{F}^A \mathcal{F}^B \tilde{\eta}_{AB} = - \mathcal{F}\overline{\mathcal{F}} \ . 
\end{equation}
We introduce the world-sheet Zweibeine $e_{\alpha}{}^{\sf a}$, with ${\sf a} = 0,1$, such that the world-sheet metric is 
\begin{equation}
h_{\alpha\beta} = e_{\alpha}{}^{\sf a}e_{\beta}{}^{\sf b} \eta_{{\sf a}{\sf b}} \ , 
\end{equation}
and we define their light-cone combinations
\begin{equation}
e_{\alpha} \equiv e_{\alpha}{}^{0} + e_{\alpha}{}^{1} \ , \qquad\qquad
\overline{e}_{\alpha} \equiv e_{\alpha}{}^{0} - e_{\alpha}{}^{1} \ , 
\end{equation}
and the world-sheet metric is 
\begin{equation}
h_{\alpha\beta} = - e_{(\alpha}\overline{e}_{\beta)} \ . 
\end{equation}
Then we introduce the light-cone combinations of $\tau_{\mu}{}^A$ as 
\begin{equation}
\tau_{\mu} \equiv \tau_{\mu}{}^0 + \tau_{\mu}{}^1 \ , \qquad\qquad
\overline{\tau}_{\mu} \equiv \tau_{\mu}{}^0 - \tau_{\mu}{}^1 \ .
\end{equation}
By using the expression for $\mathcal{F}^A$ in eqn.  (\ref{F}), we find that 
\begin{equation}
\mathcal{F} = \tau_{\mu} \partial_0 X^{\mu} - \frac{\sqrt{- h} + h_{01}}{h_{11}} \tau_{\mu}\partial_1 X^{\mu} \ , \qquad\quad
\overline{\mathcal{F}} = \overline{\tau}_{\mu} \partial_0 X^{\mu} + \frac{\sqrt{- h} - h_{01}}{h_{11}} \overline{\tau}_{\mu}\partial_1 X^{\mu} \ .
\end{equation}
At this stage one can substitute the world-sheet metric components $h_{\alpha\beta}$ in terms of the Zweibeine, and further simplify the expression above.  First, we rewrite $\sqrt{- h}$ as
\begin{equation}
\sqrt{- h} = \sqrt{(h_{01})^2 - h_{00}h_{11}} = \frac{1}{2}\sqrt{(e_0 \overline{e}_1 - e_1\overline{e}_0)^2} = \frac{1}{2} (e_1 \overline{e}_0 - e_0\overline{e}_1) \ , 
\end{equation}
where in the last equality we had to assume the positivity constraint
\begin{equation}
\label{positivity_Zweibeine}
e_1 \overline{e}_0 - e_0\overline{e}_1 \geq 0 \ .
\end{equation}  
Under this assumption, we find that 
\begin{equation}
\frac{\sqrt{- h} + h_{01}}{h_{11}} = \frac{e_0}{e_1} \ , \qquad\qquad
\frac{\sqrt{- h} - h_{01}}{h_{11}} = - \frac{\overline{e}_0}{\overline{e}_1} \ , 
\end{equation}
and therefore $\mathcal{F}$ and $\overline{\mathcal{F}}$ become
\begin{eqnarray}
\notag
\mathcal{F} = - \frac{1}{e_1} \varepsilon^{\alpha\beta} \tau_{\mu} e_{\alpha} \partial_{\beta} X^{\mu} \equiv - \frac{1}{e_1} \mathfrak{F}  \ , &&\qquad\quad
\overline{\mathcal{F}} = - \frac{1}{\overline{e}_1} \varepsilon^{\alpha\beta} \overline{\tau}_{\mu} \overline{e}_{\alpha} \partial_{\beta} X^{\mu} \equiv - \frac{1}{\overline{e}_1} \overline{\mathfrak{F}}\ , \\
&&\hspace{-1cm}\mathcal{F}^2 = - \frac{1}{h_{11}} \mathfrak{F}\overline{\mathfrak{F}} \ .
\end{eqnarray}
The divergent part of the action (\ref{action_F2}) rewrites as 
\begin{equation}
\mathcal{L}_{div} = - \frac{T c^2}{2} \sqrt{-h} h^{00} \mathcal{F}^2 = c^2 k  \, \mathfrak{F}\overline{\mathfrak{F}} \ , \qquad\qquad 
k \equiv \frac{T}{2}  \sqrt{-h} \frac{h^{00}}{h_{11}} \ . 
\end{equation}
which can be further rewritten in terms of Lagrange multipliers $(\la, \lb)$ as
\begin{equation}
\label{Ldiv_Eric}
\mathcal{L}_{div} = \la\, \mathfrak{F} + \lb\, \overline{\mathfrak{F}} - \frac{1}{c^2 k} \la \lb \ . 
\end{equation}
Finally,  in the limit $c\rightarrow\infty$ the action (\ref{action_F2}), with $\mathcal{L}_{div} $ written as in (\ref{Ldiv_Eric}), simplifies to
\begin{equation}
\label{NR_action_Eric}
S^{NR} = - \frac{T}{2} \int \dd^2 \sigma \, \bigg( \gamma^{\alpha\beta}\partial_{\alpha} X^{\mu} \partial_{\beta} X^{\nu} H_{\mu\nu} + \varepsilon^{\alpha\beta} (\lambda e_{\alpha} \tau_{\mu} + \lb \eb_{\alpha} \tb_{\mu} )\partial_{\beta}X^{\mu}  \bigg) \ . 
\end{equation}
To summarise,  the non-relativistic actions (\ref{NR_action}) and (\ref{NR_action_Eric}) are equivalent only if the world-sheet Zweibeine satisfy the positivity condition (\ref{positivity_Zweibeine}). 
If instead of choosing the positivity constraint (\ref{positivity_Zweibeine}) we would have chosen the quantity $e_1 \overline{e}_0 - e_0\overline{e}_1$ to be negative, then one would have arrived to an action which looks like (\ref{NR_action_Eric}) but with $e_{\alpha}$ and $\eb_{\alpha}$ swapped.

\subsection{$\mathbb{Z}_2$ orbifold symmetry}
\label{sec:Z2orbifold}

The non-relativistic action written in the form (\ref{NR_action_Eric}) is particularly suggestive because, as we are going to explain,  it renders manifest a  $\mathbb{Z}_2$ symmetry which will be used to take an orbifold compactification of the theory. 

The non-relativistic action in the form of eqn.  (\ref{NR_action}) is manifestly invariant under the transformation of the world-sheet Zweibeine 
\begin{equation}
\label{Zweibeine_swap}
e_{\alpha} \rightarrow \eb_{\alpha} \ , \qquad\qquad 
\eb_{\alpha} \rightarrow e_{\alpha} \ , 
\end{equation}
however the non-relativistic action in (\ref{NR_action_Eric}) is not.  A possible simple explanation for this is because under the transformation (\ref{Zweibeine_swap}) the quantity $e_1 \overline{e}_0 - e_0\overline{e}_1$ inverts sign, and therefore the positivity constraint (\ref{positivity_Zweibeine}) is no longer fulfilled.   

However, a way to make the transformation (\ref{Zweibeine_swap}) to be a symmetry of the non-relativistic action (\ref{NR_action_Eric}) is to transform simultaneously also $\la, \lb, \ta_{\mu}, \tb_{\mu}$. In particular, the transformation 
\begin{equation}
\label{Polyakov_symmetry}
e_{\alpha} \rightarrow \eb_{\alpha} \ , \qquad
\la \rightarrow \lb \ , \qquad
\ta_{\mu} \rightarrow \tb_{\mu} \ , \qquad
\eb_{\alpha} \rightarrow e_{\alpha} \ ,  \qquad
\lb \rightarrow \la \ , \qquad
\tb_{\mu} \rightarrow \ta_{\mu} \ , 
\end{equation}
leaves the action (\ref{NR_action_Eric}) manifestly invariant. 

Now, consider the following parity transformation on the $z_1$ coordinate
\begin{equation}
\label{z1_symmetry}
z_1 \rightarrow - z_1 \ . 
\end{equation}
This transformation is clearly a symmetry of the non-relativistic action in the Nambu-Goto form (\ref{NR_action_NambuGoto}),  since the action is solely expressed in terms of the AdS$_5\times$S$^5$ string Newton-Cartan data $\{\tau_{\mu}{}^A, m_{\mu}{}^A, e_{\mu}{}^a\}$, which depends quadratically on the coordinate $z_1$. 

Consider now this parity transformation on $z_1$ for the Polyakov action.  The kinetic term that couples to $H_{\mu\nu}$ is manifestly invariant, but the Lagrange multipliers term is not. However, we observe that since $\tau_{\mu}{}^A$ is diagonal,  inverting the sign of $z_1$ can be equivalently seen as transforming $\ta_{\mu} \rightarrow \tb_{\mu}$ and $\tb_{\mu} \rightarrow \ta_{\mu}$,  e.g.
\begin{equation}
\ta_{\mu} \partial_{\alpha}X^{\mu} = \tau_{t}{}^0  \partial_{\alpha} t + \tau_{z_1}{}^1 \partial_{\alpha} z_1 \qquad\rightarrow \qquad
 \tau_{t}{}^0  \partial_{\alpha} t + \tau_{z_1}{}^1 \partial_{\alpha} (-z_1) = \tb_{\mu} \partial_{\alpha}X^{\mu} \ . 
\end{equation}
Therefore, by recalling that the Polyakov non-relativistic action enjoys the invariance (\ref{Polyakov_symmetry}), we have that the simultaneous transformation 
\begin{eqnarray}
z_1 \rightarrow - z_1 \ , \quad
e_{\alpha} \rightarrow (-1)^{s}\, \eb_{\alpha} \ , \quad
\la \rightarrow (-1)^{s}\, \lb \ , \quad
\eb_{\alpha} \rightarrow (-1)^{s} e_{\alpha} \ ,  \quad
\lb \rightarrow (-1)^{s} \la \ , 
\end{eqnarray}
where $s$ is either $0$ or $1$,  leaves the Polyakov non-relativistic action invariant.  This replaces the simple notion of parity symmetry (\ref{z1_symmetry}) on the $z_1$ coordinate for the Polyakov action.  In section (\ref{sec:twisted_states}) we shall use this $\mathbb{Z}_2$ symmetry to make an orbifold compactification of the theory on the $z_1$ coordinate, which will allow us to construct closed string solutions with twisted boundary conditions.

\subsection{Conformal gauge and equations of motion}

In this section we shall study the consequences of fixing conformal gauge in the non-relativistic action (\ref{NR_action_Eric}). As we keep the world-sheet relativistic, we are allowed to fix the word-sheet metric in the same fashion as in a relativistic string theory. In particular, we fix it to the conformal gauge, i.e. $h_{\alpha\beta} = \eta_{\alpha\beta}$.

We should start by commenting on the remnant gauge redundancy. Fixing conformal gauge in the relativistic Polyakov action leaves a Diff$_+ \oplus$Diff$_-$ residual gauge symmetry that allows for separate diffeomorphisms of the world-sheet light-cone coordinates. Despite the $\lambda_A$ being scalars under world-sheet diffeomorphisms, they acquire non-trivial transformation properties when we fix conformal gauge. In particular, their light-cone combinations $\lambda_1 \pm \lambda_0$ transform as world-sheet $1$-forms \cite{Bergshoeff:2018yvt}. This implies that the non-relativistic action in conformal gauge also has a residual Diff$_+ \oplus$Diff$_-$ gauge symmetry.

In terms of the world-sheet Zweibeine $e_{\alpha}{}^{\sf a}$, conformal gauge means that they are $\mathfrak{so}(1,1)$ matrices.  This leaves us with enough freedom to make the following choice
\begin{equation}
e_{\alpha}{}^{\sf a} = \begin{pmatrix}
-1 & 0 \\
0 & -1
\end{pmatrix}\qquad\Longrightarrow\qquad
e_{\alpha} = ( -1, -1) \ , \qquad
\eb_{\alpha} = ( -1,  1) \ , 
\end{equation}
which satisfies the positivity constraint (\ref{positivity_Zweibeine}). Therefore, the two actions (\ref{NR_action}) and (\ref{NR_action_Eric}) are equivalent.  The relation between $(\lambda, \lb)$ and $(\lambda_0, \lambda_1)$ which makes the two actions equivalent within this choice of Zweibeine is
 \begin{equation}
 \lambda = \frac{1}{2} (\lambda_0 + \lambda_1) \ , \qquad\qquad
 \lb = \frac{1}{2} (\lambda_1 - \lambda_0) \ . 
 \end{equation}

In order to derive the equations of motion for the fields $X^{\mu}, \la, \lb$ from the action (\ref{NR_action_Eric}) by using the variational principle, we need to demand that the fields are fixed on the Cauchy surfaces at initial and final time, i.e. 
$\delta X^{\mu}(\tau_i, \sigma) =  \delta X^{\mu}(\tau_f, \sigma) = 0$,  
for every value of $\sigma$, and the same condition for $\la$ and $\lb$.  This is not enough to set the surface term completely to zero,  but we remain with 
\begin{equation}
\label{surface_term}
\delta S^{NR}_{\text{surface}} = -\frac{T}{2} \left[ 2 X'^{\mu} H_{\mu\nu} \delta X^{\nu} - (\la \tau_{\mu} + \lb  \tb_{\mu}) \delta X^{\mu} \right]_{\sigma=0}^{\sigma=2 \pi} \  ,
\end{equation}
where conformal gauge has already been implemented. This remaining surface term will be set to zero by taking opportune boundary conditions for our solutions (closed and twisted strings).  

After eliminating all surface terms, the equations of motion for the fields $X^{\mu}, \lambda_0, \lambda_1$ obtained from the variation of the non-relativistic action in conformal gauge can be written as
\begin{align}
	&\mathcal{E}_t \equiv \frac{\dd}{\dd\tau} \frac{\partial \mathcal{L}}{\partial \dot{t}} + \frac{\dd}{\dd\sigma} \frac{\partial \mathcal{L}}{\partial t'}=0 \ ,  \label{EoM1} \\
	&\mathcal{E}_{z_1} \equiv \frac{\dd}{\dd\tau} \frac{\partial \mathcal{L}}{\partial \dot{z}_1} + \frac{\dd}{\dd\sigma} \frac{\partial \mathcal{L}}{\partial z'_1} - \frac{\partial \mathcal{L}}{\partial z_1}=0 \ ,  \label{EoM2} \\
	&\mathcal{E}_{z_m} \equiv \frac{1+ (\frac{z_1}{2R})^2}{R^2(1- (\frac{z_1}{2R})^2)} z_m (-\dot{t}^2+t^{\prime 2}) -  \frac{z_m}{2R^2(1- (\frac{z_1}{2R})^2)} (-\dot{z}_1^2+z_1^{\prime 2}) \notag \\
	&\hspace{1cm}+  \frac{z_1}{R^2(1- (\frac{z_1}{2R})^2)} (-\dot{z}_1\dot{z}_m+z'_1 z'_m) -\ddot{z}_m+z''_m  =0 \ , \label{EoM3} \\
	&\mathcal{E}_{y_i} \equiv \ddot{y}_i - y''_i=0 \ ,\\
	&\mathcal{E}_{\phi} \equiv \ddot{\phi} - \phi''=0 \ ,\\
	&\mathcal{E}_{\lambda_0} \equiv \mathcal{F}^0=- \frac{1+ (\frac{z_1}{2R})^2}{1- (\frac{z_1}{2R})^2} \left( \dot{t} + \frac{z'_1}{1 + (\frac{z_1}{2R})^2} \right)=0 \ , \\
	&\mathcal{E}_{\lambda_1} \equiv \mathcal{F}^1 =\frac{1+ (\frac{z_1}{2R})^2}{1- (\frac{z_1}{2R})^2} \left( t' + \frac{\dot{z}_1}{1 + (\frac{z_1}{2R})^2} \right)=0 \ .\label{EoM4}
\end{align}
We do not write explicitly the equations of motion for the $t$ and $z_1$ coordinates due to their length and intricacy.

In addition to the equations of motion, fixing conformal gauge forces us to also impose the Virasoro constraints. In the conformal gauge, they can be written as
\begin{align}
	V_1&=H_{\mu \nu} \dot{X}^\mu X^{\prime \nu}+\frac{1}{2} \lambda_A \tau^A_\nu X^{\prime \nu}= 0 \ , \label{VC1} \\
	V_2&=H_{\mu \nu} \dot{X}^\mu \dot{X}^\nu + H_{\mu \nu}  X^{\prime \mu} X^{\prime \nu} + \lambda_0 \tau^1_\mu X^{\prime \mu} +\lambda_1 \tau^0_\mu X^{\prime \mu}=0 \label{VC2} \ .
\end{align}
Notice that the terms involving the Lagrange multipliers are never accompanied by a $\dot{X}$ factor. As we will always have a target-space coordinate that depends on $\sigma$ due to the constraints imposed by said Lagrange multipliers, they will always contribute to the Virasoro constraints.

\vspace{3mm}
\noindent\textbf{Equations of motion for $\lambda_{A}$.} Let us come back and analyse the equations of motion (constraints) $\mathcal{F}^A=0$ in more depth. It is immediate to see that they can be simplified if we perform the change of variables $z_1=2 R \tan [Z/(2R)]$, giving us
\begin{equation}
	\dot{t}+Z'=t'+\dot{Z}=0 \ . \label{constraints}
\end{equation}
The two constraints become nothing but the analytical continuation of the Cauchy-Riemann equations after the transformations $t\rightarrow it$ and $\tau \rightarrow i\tau$. This indicates us that the functions $t$ and $Z$ satisfy a wave equation. For concreteness, let us say that
\begin{equation}
\label{t_z1_f_g}
	t= 2R \bigg(g(\tau - \sigma) - f(\tau + \sigma)\bigg)  \ ,  \qquad \qquad z_1= 2 R \tan [f(\tau + \sigma) + g(\tau - \sigma) + c ] \ ,
\end{equation}
where $c$ is an arbitrary constant and $f,g$ are arbitrary functions. Although not obvious at first sight, we can further simplify these solutions. As we commented above, the conformal gauge does not completely fix all the gauge freedom of the theory. We can use the residual Diff$_+ \oplus$Diff$_-$ symmetry to set the functions $f$ and $g$ to be linear in their arguments and with opposite coefficients,  such that (\ref{t_z1_f_g}) becomes
\begin{equation}
\label{diff+diff-fixed}
t=\kappa \tau\ ,  \qquad \qquad z_1= 2 R \tan \bigg[ - \frac{\kappa}{2R} (\sigma-\sigma_0) \bigg] \ .
\end{equation}
Notice that $\sigma_0$ is just a rewriting of the constant $c$. Fixing these two coordinates is enough to exhaust the residual gauge freedom of conformal gauge, so we do not have to worry about additional issues from gauge freedom.

We want to stress that this is not the only way to fix the Diff$_+ \oplus$Diff$_-$ residual symmetry. Another possible gauge fixing we will be using would be to set the Lagrange multipliers $\lambda_A$ to a constant. We will mostly consider here the choice $t=\kappa \tau$ as is a more intuitive and natural, but we say some words about this other choice in the appendix.

Although the shared radius of the AdS and sphere can be set to one by an appropriate field redefinition, it is worth saying some words about the large $R$ limit before doing so. If we take the naïve $R\rightarrow \infty$ limit, we get
\begin{equation}
\label{largeR_linear}
	t=\kappa \tau \ , \qquad z_1= -\kappa (\sigma-\sigma_0) - \frac{\kappa^3}{12 R^2} (\sigma-\sigma_0)^3 + \mathcal{O} (R^{-4}) \ ,
\end{equation}
that is, the tangent becomes just a linear contribution plus large radius correction. However, this result is not correct due to the closed string condition. Demanding $z_1(\sigma + 2\pi) = z_1 (\sigma)$ restricts $\kappa/R$ to be an integer number. Thus, we have to consider $\kappa$ of the same order of $R$ when we compute the limit. One way to eliminate this restriction is to consider instead \emph{twisted} closed string boundary conditions, allowed in the $\mathbb{Z}_2$ orbifold compactification of the theory, where $\kappa/R$ is no longer restricted to be an integer.  We will discuss this in section \ref{sec:twisted_states}.

It is a remarkable fact that the tangent dependence of $z_1$ on $\sigma$ in (\ref{diff+diff-fixed}) makes the string wrapping around the non-compact $z_1$ coordinate. As later described, when the closed string boundary conditions are imposed, the string goes along $z_1$ up to $+ \infty$ and comes back from $-\infty$ several times. However, the coordinate $Z = 2R \arctan[z_1 / (2R)]$ is compact, due to the property of the $\arctan$. Therefore the string winding the $z_1$ coordinates through the infinities can be mapped into winding of the compact coordinate $Z$. This should be then in agreement with the Gomis-Ooguri result for NR strings in flat spacetime \cite{Gomis:2000bd}, which in order to have a non-trivial spectrum they need to wrap around a longitudinal compact spatial coordinate. 
Instead, when the twisted boundary conditions are imposed, the string wraps the $z_1$ direction around the orbifold fixed point $z_1 = 0$.

To end this section, we want to comment that the non-relativistic action (\ref{NR_action}) is invariant under shifts of the $t$ coordinate and rotations between the $z_2$, $z_3$ and $z_4$ coordinates.  This is just a consequence of the symmetries of the AdS$_5\times$S$^5$ string Newton-Cartan data. 
In addition, the directions formerly associated to $S^5$ in the relativistic theory, i.e. $(\phi, y_i)$,  now become flat directions after taking the non-relativistic limit.  Thus, we can define the following conserved quantities
\begin{align}
	E&= - \frac{T}{2} \int_0^{2\pi} d\sigma \frac{1+ (\frac{z_1}{2R})^2}{1- (\frac{z_1}{2R})^2} \left( \frac{2 ( z_2^2 +z_3^2 + z_4^2) \dot{t} }{\left( 1- (\frac{z_1}{2R})^2 \right) ^2 } -\lambda_0 \right) \ ,\label{classicalE} \\
	S_{m n} &= - \frac{T}{2} \int_0^{2\pi} d\sigma \frac{z_m \dot{z}_n - z_n \dot{z}_m }{(1- (\frac{z_1}{2R})^2)^2} \ , \\
	p_{i} &= - \frac{T}{2} \int_0^{2\pi} d\sigma \dot{y}_i \ , \phantom{0000000000000} J= - \frac{T}{2} \int_0^{2\pi} d\sigma \dot{\phi} \ .
\end{align}
Since $\phi$ becomes non-compact in the NR limit, the conserved charge $J$ has not meaning of angular momentum as in the relativistic case but as a linear momentum. Nevertheless we shall call it $J$ to remind similarities with the relativistic BMN dispersion relation. 
This set of conserved quantities are not exhausting the whole set of Noether charges associated with the global symmetries of the theory, but are the ones of interest for the classical string solutions we will present.

\section{Classical NR string solutions in conformal gauge}
\label{sec:clas_sol}
\setcounter{equation}{0}

In this section we will study some particular solutions of the equations of motion (\ref{EoM1})-(\ref{EoM4}).  In particular, we are able to find classical configurations reminiscent of the BMN string and GPK string that appear in relativistic AdS$_5\times$S$^5$. 

\subsection{Closed NR string sector}

We shall assume that $\sigma$ ranges from $0$ to $2 \pi$,  and we shall impose that the fields $X^{\mu}$ and $\lambda_A$ satisfy the periodicity condition
\begin{equation}
\label{closed_bc}
X^{\mu}(\tau, \sigma) = X^{\mu}(\tau, \sigma + 2\pi) \ , \qquad\qquad
\lambda_A(\tau, \sigma) = \lambda_A(\tau, \sigma + 2\pi) \ . 
\end{equation}
For simplicity,  here in the closed string sector we set $R=1$, but we will leave it generic when we will study the twisted sector. 

\subsubsection{BMN-like string}
\label{sec:BMN_like_string}

Let us start by proposing the ansatz
\begin{align*}
	t&=\kappa \tau \ , & z_1 &=z_1(\sigma) \ , & z_m &=0 \ , & y_i &=0 \ , & \phi &=\phi (\tau , \sigma) \ , & \lambda_A = \lambda_A(\tau,\sigma) \ .
\end{align*}
After substituting this ansatz into the equations of motion~(\ref{EoM1}) and (\ref{EoM2}) and the constraint $\mathcal{F}^0$, we find\footnote{Because the constraints enforced by our Lagrange multipliers involve the generalised velocities instead on only constraining the generalised coordinates, we get different results depending on if we substitute them at the level of the Lagrangian or at the level of the equations of motion. As our ansatz enforces one of the two constraints, we have to be careful on when to substitute it. The correct way of doing it in our case is to make the substitution at the level of the equations of motion, which keeps the dependence on $\lambda_1$ alive.}
\begin{align}
		&\lambda'_1 - \dot{\lambda}_0+\frac{z_1}{1- (\frac{z_1}{2})^4} \lambda_1 z'_1=0 \ , \\
	&\lambda'_0 - \dot{\lambda}_1 + \frac{z_1}{1- (\frac{z_1}{2})^2} \lambda_1 \kappa=0 \ , \\
	&\kappa + \frac{z'_1}{1 + (\frac{z_1}{2})^2}=0 \ , \\
	&\ddot{\phi} - \phi''=0 \ ,
\end{align}
while the remaining equations of motion and the remaining constraint identically vanish. These equations are satisfied if we choose
\begin{align}
\label{BMN_z1_phi}
	z_1 &= 2\tan \left[ -\frac{\kappa}{2} (\sigma-\sigma_0) \right] \ , & \phi &=\phi_+ (\tau+\sigma) + \phi_- (\tau-\sigma) \ , \\
	\label{BMN_lambda}
	\lambda_1 + \lambda_0 &= \Lambda_+ (\tau+\sigma) \cos \left[ \kappa (\sigma-\sigma_0) \right] \ , &
	\lambda_1 - \lambda_0 &= \Lambda_-(\tau-\sigma) \cos \left[ \kappa (\sigma-\sigma_0) \right] \ , 
\end{align}
where $\Lambda_\pm$ and $\phi_\pm$ are arbitrary functions not fixed by the equations of motion. By demanding our solution to be a closed string, the condition e.g. $z_1 (\sigma + 2\pi)=z_1 (\sigma )$, among the others, forces $\kappa$ to be an integer.  We can understand this solution as a string winding repeatedly around $\mathbb{R}$, going to infinity and coming back from minus infinity. Thus, the solution we are considering has the topology of the one-point compactification of $\mathbb{R}$ instead of the topology of $\mathbb{R}$.

Let us now move our attention to the Virasoro constraints (\ref{VC1}) and (\ref{VC2}). For our ansatz, they take the form
\begin{align}
	V_1&\propto \frac{\lambda_1 z^{\prime}_1}{1 - (\frac{z_1}{2})^2} -2 \dot{\phi} \phi' \ , \\
	V_2&\propto - \frac{\lambda_0 z^{\prime}_1}{1 - (\frac{z_1}{2})^2} +(\dot{\phi})^2 + (\phi')^2 \ ,
\end{align}
If we now substitute the solutions we found for $z_1$ and the Lagrange multipliers eqns. (\ref{BMN_z1_phi}) and (\ref{BMN_lambda}), they become algebraic equations for $\Lambda_\pm$
\begin{gather}
	\kappa \Lambda_+ =(\dot{\phi})^2 + (\phi')^2 +2 \dot{\phi} \phi'=(\dot{\phi}+\phi')^2=(\dot{\phi}_+ +\phi'_+ )^2 \ , \\
	\kappa \Lambda_- =-(\dot{\phi})^2 - (\phi')^2 +2 \dot{\phi} \phi'=-(\dot{\phi}-\phi')^2=-(\dot{\phi}_- -\phi'_- )^2 \ .
\end{gather}
Notice that these algebraic equations are consistent with the results of our differential equation, in the sense that $\Lambda_+$ only depends on $\tau+\sigma$ and similarly $\Lambda_-$ only depends on $\tau-\sigma$. In fact, we could instead had solved the Virasoro constraints first and then substituted the results into the equations of motion for $t$ and $z_1$. In that case we would have obtained that these equations of motion are just the wave equation for $\phi$, indicating us that we do not have to impose any further restriction to the functions $\phi_\pm$.

At this point we need to consider the periodicity on $\lambda_A$.  Since the Virasoro constraints relate $\Lambda_{\pm}$ with $\phi$, imposing $\sigma$-periodicity on $\phi$ and $z_1$ is enough to guarantee periodicity on $\lambda_A$.

Although any $\sigma$-periodic function $\phi$ that satisfies the wave equation is allowed, there is a particular case we would like to look into in more detail: the case $\phi=w \tau$, with $w$ constant. This case can be seen as a generalization of the BMN string we discussed in relativistic AdS$_5\times$S$^5$, described by the trajectory $t=\phi=\kappa \tau$. In the non-relativistic case the solution no longer describes a point-like particle but an extended object in the $z_1$ direction. It is easy to see that the dispersion relation of our BMN-like string is
\begin{equation}
\label{BMN_dispersion}
	J=-\pi w T \ , \qquad E=-\frac{T}{2} \, \frac{2\pi w^2}{\kappa}=\frac{-J^2}{\pi \kappa T} \ .
\end{equation}
Despite the similarity of this expression with the relativistic BMN string, there are two key differences we should emphasize. First, the relation $E\propto \kappa T$, characteristic of relativistic BMN string, no longer holds here. This happens because the $H_{tt}$ element of the metric vanishes when we turn off all the other fields instead of becoming $-1$. This behaviour hinders the applications of most of the usual methods developed for relativistic string theory, as can be seen in \cite{Fontanella:2021hcb}. The second difference we want to point out is that the constants $w$ and $\kappa$ are no longer related. Originally, the Virasoro constraints forced them to be equal, but in this setting they fix the Lagrange multipliers instead. In fact, another way to write the dispersion relation (\ref{BMN_dispersion}) is 
\begin{equation}
E - J = \left(\frac{w}{\kappa} - 1\right) J \ , 
\end{equation}
which reduces to the relativistic BMN dispersion relation when $w = \kappa$.

\subsubsection{Generalized BMN-like string}

Here we will generalize the previous BMN-like solution, by giving non-trivial dependence on $\sigma$ to the other coordinates in AdS$_5$, i.e.
\begin{align*}
	t&=\kappa \tau \ , & z_1 &=2\tan \left[ -\frac{\kappa}{2} (\sigma-\sigma_0) \right] \ , & z_m &=z_m(\sigma) \ , & y_i &=0 \ , & \phi &=\phi (\tau , \sigma) \ , & \lambda_A = \lambda_A(\tau,\sigma) \ .
\end{align*}

First, we will focus our attention on the equation of motion for the $z_m$ coordinates. Thanks to the $SO(3)$ symmetry that rotates the $z_m$ coordinates, these three equations of motion are the same, so we can analyse them all together at the same time. Instead of considering directly (\ref{EoM3}), we can bring that equation to a simpler form if we write it as a function of $z_1$ instead of a function of $\sigma$. If introduce the variable $x=z_1/2$ and use the constraint from the Lagrange multiplier to write $z'_1$ and $z^{\prime \prime}_1$ in terms of $z_1$, we get
\begin{equation}
	\frac{d^2 z_m}{d x^2}+ \left( \frac{1}{x-i}+\frac{1}{x+i} -\frac{2}{x-1}-\frac{2}{x+1} \right) \frac{d z_m}{d x}+ \frac{2(3 +x^2)}{x^4-1}z_m=0 \ .
\end{equation}
This differential equation falls into the category of \emph{generalized Lamé equations}, and it can be solved in terms of simple functions if the numerator in the $z_m$ coefficient has the appropriate form.\footnote{In particular, a generalized Lamé equation has polynomial solutions, called ``Heine–Stieltjes polynomials'', if the numerator in the $z_m$ coefficient is a Van Vleck polynomial. The zeroes of the Heine–Stieltjes polynomials fulfil the Bethe Ansatz equations for the Gaudin Model\cite{Gaudin}, while the zero of the Van Vleck polynomials fulfil an equation that resembles an auxiliary Bethe equation (as it depends also on the positions of the zeros of the Heine–Stieltjes polynomials). We refer the reader to \cite{VanVleck} for more information. If a Lax connection for NRST is found, it would be interesting to see if such ``Bethe equation lookalikes'' could be obtained from first principles from it.} Luckily, that is our case, and the solution is given by
\begin{equation}
\label{zmsolution_x_dep}
z_m=C^1_m x +C^2_m \left[ 1+4 x \arctan (x) -x^2 \right] \ ,
\end{equation}
or equivalently in terms of $\sigma$ as
\begin{eqnarray}
\notag
z_m&=&C^1_m \tan \left[ -\frac{\kappa}{2} (\sigma-\sigma_0) \right] \\
&+&C^2_m \left\{ 1 -2 \kappa (\sigma-\sigma_0) \tan \left[ -\frac{\kappa}{2} (\sigma-\sigma_0) \right]  - \tan^2 \left[ -\frac{\kappa}{2} (\sigma-\sigma_0) \right] \right\} \ , \label{zmsolution}
\end{eqnarray}
where $C^1_m$ and $C^2_m$ are integrations constants. Notice that we have used the relation $\arctan (\tan (x))=x$ to get this expression, i.e., we have chosen the branches of the arctangent in such a way that this combination gives us a continuous function.\footnote{Any different choice of branch can be reduced to our choice by opportunely redefining the constant $C^1_m$. We thank R. Ruiz for pointing this out.}

At this point, we should check if the solution we have found fulfils the equation of motion associated to $z_1$, $t$, and the Virasoro constraints. The first one can be written as
\begin{multline}
	z_m z_m z^{\prime \prime}_1+\left( 2 z_m z'_m -\frac{3z_1 z_m z_m}{2\left[ 1-\left( \frac{z_1}{2}\right)^2 \right]} \right) z'_1 +  2\left[ 1-\left( \frac{z_1}{2}\right)^2 \right] \kappa \lambda_0 \\
	-\left[ 1-\left( \frac{z_1}{2}\right)^2 \right]^2 \lambda'_0 +\left[ 1-\left( \frac{z_1}{2}\right)^2 \right]^2 \dot{\lambda}_1 + \left(2 z'_m z'_m + \frac{2\left[ 2+\left( \frac{z_1}{2}\right)^2 \right] \kappa^2 z_m z_m}{1-\left( \frac{z_1}{2}\right)^2}\right) z_1=0 \ ,
\end{multline}
Substituting the explicit expressions for $z_1$ and $z_m$, the equation becomes a partial differential equation for the Lagrange multipliers. This equation is difficult to solve, but we can find simpler equations for the sum of the Lagrange multipliers by adding the equation of motion for $t$. After substituting the ansatz, $\mathcal{E}_t$ takes the form
\begin{equation}
	\kappa \tan (\kappa \sigma) \lambda_1 + \lambda'_1 - \dot{\lambda}_0=0
\end{equation}
We can check that the solution of the differential equation $\mathcal{E}_{z_1}+\mathcal{E}_t=0$ can be written as
\begin{eqnarray}
\notag
	\lambda_1+\lambda_0&=& C^3_+ \cos [ \kappa (\sigma-\sigma_0)] -\frac{\kappa}{4} \bigg( C^1_m C^1_m +4\kappa (\sigma-\sigma_0)  C^1_m C^2_m \\
	&+&[4+4\kappa^2 (\sigma-\sigma_0)^2+\sec^2 (\kappa (\sigma-\sigma_0)/2) ] C^2_m C^2_m \bigg) \ ,
\end{eqnarray}
where $C^3_+$ is a generic function of $\tau+\sigma$. Similarly, if we instead look at the combination $\mathcal{E}_{z_1}-\mathcal{E}_t=0$, we get a differential equation whose solution is given by
\begin{eqnarray}
\notag
	\lambda_1-\lambda_0 &=& C^3_- \cos [ \kappa (\sigma-\sigma_0)] +\frac{\kappa}{4} \bigg( C^1_m C^1_m +4\kappa (\sigma-\sigma_0) C^1_m C^2_m \\
	&+&[4+4\kappa^2 (\sigma-\sigma_0)^2+\sec^2 (\kappa (\sigma-\sigma_0)/2) ] C^2_m C^2_m \bigg) \ ,
\end{eqnarray}
where $C^3_-$ is a generic function of $\tau-\sigma$. 

Let us now move to the Virasoro constraints. For our problem at hand, they reduce to
\begin{align}
	V_1&\propto \frac{\lambda_1 z^{\prime}_1}{1 - (\frac{z_1}{2})^2} -2 \dot{\phi} \phi' \ , \\
	V_2&\propto \frac{z^{\prime 2}_m}{(1- (\frac{z_1}{2})^2)^2} + \frac{z_m^2 z^{\prime 2}_1 }{2(1- (\frac{z_1}{2})^2)^3} - \kappa^2 \frac{(1+ (\frac{z_1}{2})^2)z_m z_m}{(1- (\frac{z_1}{2})^2)^3}- \frac{\lambda_0 z^{\prime}_1}{1 - (\frac{z_1}{2})^2} +(\dot{\phi})^2 + (\phi')^2 \ .
\end{align}
Similarly to the previous case, we can linearly combine these two equations in order to get algebraic equations for $\lambda_1 \pm \lambda_0$, so we can fix the integration constants $C^3_\pm$ in terms of derivatives of $\phi$. In particular, we find
\begin{equation}
	\kappa C^3_-=(\dot{\phi} -\phi')^2 - \kappa^2 C^2_m C^2_m \ , \qquad \kappa C^3_+= \kappa^2 C^2_m C^2_m - (\dot{\phi} +\phi')^2 \ .
\end{equation}
Thus, the Virasoro constraints again just fix the integration constants unfixed by the equations of motion of $t$ and $z_1$, and they impose no further restrictions to the solution.

The only remaining step is to impose the closed string conditions to the coordinates and the Lagrange multipliers. First, the conditions $z_m (\sigma +2\pi) = z_m (\sigma)$ forces us to set $C^2_m=0$ due to the presence of the linear and quadratic terms in $\sigma$ in the solution associated to $C^2_m$. This is again a consequence of our choice of branch of the arctangent, where $\arctan (\tan(x))=x$ for any real value of $x$. Regarding the Lagrange multipliers, their periodicity in the $\sigma$ coordinate is assured provided we enforce periodicity on the $\phi$ coordinate.

As we are forced to set $C^2_m=0$, all the $z_m$ coordinates are proportional to $z_1$. We might be tempted to understand this solutions as an $SO(4)$ rotation of the previous BMN-like string, but this is not completely right. The NRST we are considering breaks the $SO(4)\subset SO(2,4)$ symmetry of AdS$_5$ to an $SO(3)$ symmetry that does not involve the $z_1$ coordinate.

Similarly to the previous solution we studied, we want to end this section by looking at the conserved charges associated to this solution. If we focus again in the case with $\phi=w\tau$, due to its interest, the angular momentum and energy are given by
\begin{equation}
\label{genBMN_dispersion}
	J=-\pi w T \ , \qquad E= -\frac{T}{2\pi} \left( \frac{w^2}{\kappa}+ 2\pi \kappa C^2_m C^2_m \right)
\end{equation}
where we have assumed that $\kappa$ is an integer. Only the contribution from $C^2_m C^2_m$ survives because of its non-periodicity. As we have to set it to zero due to periodicity, we get the same dispersion relation we got for the previous solution.

\subsubsection{GKP-like string}

In this section we will try to add a dependence on $\tau$ to the $z_m$ coordinates by means of the $SO(3)$ symmetry they present. For classical rotating NR string solutions in flat spacetime, see \cite{Gomis:2004ht}. In order to do that, it is more convenient to write them in coordinates that make that symmetry manifest
\begin{equation}
\label{zm_profile_GKP}
	z_2=z \cos \theta \, \cos \varphi \ , \qquad z_3=z \cos \theta \,  \sin \varphi \ , \qquad z_4=z\sin \theta \ .
\end{equation}
Thanks to the fact that $H_{z_2 z_2}=H_{z_3 z_3}=H_{z_4 z_4}=(1-\tfrac{z_1^2}{4})^{-2}$, we can rewrite our metric as
\begin{equation}
	H_{z_m z_m} dz_m dz_m= H_{z_m z_m} dz^2 + H_{z_m z_m} z^2 (d \theta^2 + \cos^2 \theta d\varphi^2) \ .
\end{equation}

Drawing inspiration from the previous result, we are going to assume the following ansatz
\begin{align*}
	t&=\kappa \tau \ , & z_1 &=2\tan \left[ -\frac{\kappa}{2} (\sigma-\sigma_0) \right] \ , & z &=z(\sigma) \ , & y_i &=0 \ , & \phi &=\phi (\tau , \sigma) \ , & \lambda_A = \lambda_A(\tau,\sigma) \ ,
\end{align*}
while we will not impose any restriction on $\theta$ and $\varphi$ yet. Then, the equations of motion for these two coordinates are
\begin{align*}
	&2 H_{z_m z_m} z z' \theta ' + z^2 H'_{z_m z_m} \theta ' + z^2 H_{z_m z_m} \left[ \cos \theta \sin \theta (\varphi^{\prime 2} - \dot{\varphi}^2 ) +\theta^{\prime \prime} -\ddot{\theta} \right]=0 \ , \\
	&2 H_{z_m z_m} \cos \theta \, z z' \varphi ' + z^2 \cos \theta H'_{z_m z_m} \varphi ' + z^2 H_{z_m z_m} \left[ \cos \theta \, ( \varphi^{\prime \prime} -\ddot{\varphi} ) -2 \sin \theta \, ( \theta '\varphi ' -\dot{\theta} \dot{\varphi} )  \right]=0 \ ,
\end{align*}
where we have defined $H'_{z_m z_m}= z'_1 \partial_{z_1} H_{z_m z_m}$ to alleviate the notation. It is easy to check that $\theta=0$ and $\varphi= \omega \tau$ (and any other configuration related by an $SO(3)$ rotation) is a solution to these equations of motion. We shall focus on this solution from now on. 

Let us now check the equation of motion for $z$. After performing the same changes of variables and redefinitions we used in the previous section, we find
\begin{equation}
	z^{\prime \prime}+ \left( \frac{1}{x-i}+\frac{1}{x+i} -\frac{2}{x-1}-\frac{2}{x+1} \right) z'+ \frac{2(3 +x^2)+4\frac{\omega^2 (x^2-1)}{\kappa^2 (x^2+1)}}{x^4-1}z=0 \ .
\end{equation}
The differential equation retains most of the structure, but it has an extra terms that makes it no longer a generalized Lamé equation. Not only that, but it also ceases to be Fuchsian, as it has second order poles for $\omega \neq 0$. Nevertheless, the differential equation can still be solved
\begin{align}
	z&=\kappa \sqrt{ 16x^2 + n^2 (x^2-1)^2} \left( C^1 \sin \left[ \arctan \, x -\arctan \left( \frac{4x}{n(x^2-1)} \right) \right] \right. \notag \\
	&+\left. C^2 \cos \left[ \arctan \, x -\arctan \left( \frac{4x}{n(x^2-1)} \right) \right]\right) \ ,
\end{align}
where $C^i$ are integration constants. In addition, we have written $\omega= \frac{n\kappa}{2}$, as it seems a natural redefinition from the form of the differential equation. Interestingly, this expression can be rewritten as a polynomial in $x$ as long as $n$ is an integer. Using the properties of the trigonometric functions, we find
\begin{align}
	z&=\frac{1}{2 \sqrt{(1+x^2)^n}} \left( \left[ (1+ix)^n +(1-ix)^n \right] \left( 4 x C^1 +n (x^2-1) C^2 \right) \vphantom{\frac{1}{1}} \right. \notag \\
	&\left. +i \left[ (1+ix)^n -(1-ix)^n \right] \left( 4 x C^2 -n (x^2-1) C^1 \right) \vphantom{\frac{1}{1}} \right) \ . \label{secondrepofz}
\end{align}
Despite their appearance, these polynomials have real coefficients as long as $n$ is integer. In particular, if we rewrite $\tan [n \arctan(x)]$ as a rational function of $x$, the numerator and denominator are equal to the polynomials $(i)^n\left[ (1+ix)^n \pm(1-ix)^n \right]$ respectively
. Notice that the $\omega\rightarrow 0$ limit recovers the differential equation from the previous section, but we are not able to recover the functions accompanying $C^2_m$ in (\ref{zmsolution}). This happens because the two independent solutions of the differential equation we found become degenerate in the $\omega \rightarrow 0$ limit. This means that the second solution has to be obtained by other methods, e.g., via the Wronskian.

Let us move now our attention to the Virasoro constraints. Substituting our ansatz yields
\begin{align}
	V_1&\propto \frac{\lambda_1 z^{\prime}_1}{1 - (\frac{z_1}{2})^2} -2 \dot{\phi} \phi'=0 \ , \\
	V_2&\propto \frac{z^2 z_1^{\prime 2} -2 z^2 \left(1+\tfrac{z_1^2}{4}\right) \kappa^2}{2 \left(1-\tfrac{z_1^2}{4}\right)^3} +\frac{z^2 \omega^2 +z^{\prime 2}}{\left(1-\tfrac{z_1^2}{4}\right)^2} -\frac{\lambda_0 z'_1}{1-\tfrac{z_1^2}{4}} +(\dot{\phi})^2 + (\phi')^2 =0\ .
\end{align}
We have again two algebraic equations for the Lagrange multipliers that we can solve immediately
\begin{eqnarray}
\notag
\kappa (\lambda_1 +\lambda_0) &=&  \kappa^2 z^2 (1+\sec [\kappa (\sigma-\sigma_0)]) \frac{4-n^2 (1+\cos [ \kappa (\sigma-\sigma_0)]) }{16}\\
&-&z^{\prime 2} \sec [ \kappa (\sigma-\sigma_0)] \cos \, \frac{\kappa (\sigma-\sigma_0)}{4} - (\dot{\phi} + \phi')^2 \cos [\kappa (\sigma-\sigma_0)]\ , \\
\notag
\kappa (\lambda_1 -\lambda_0) &=&  -\kappa^2 z^2 (1+\sec [ \kappa (\sigma-\sigma_0)]) \frac{4-n^2 (1+\cos [ \kappa (\sigma-\sigma_0)]) }{16} \\
&+& z^{\prime 2} \sec [\kappa (\sigma-\sigma_0) ] \cos \frac{\kappa (\sigma-\sigma_0)}{4} + (\dot{\phi} - \phi')^2 \cos [\kappa (\sigma-\sigma_0)] \ .
\end{eqnarray}

As in the BMN-like case, when we substitute these solutions into the equations of motion for $t$ and $z_1$, we get that $\phi$ and $z$ have to satisfy their respective equations of motion. This shows that the equations of motion for $t$ and $z_1$ provide no further condition to the solution.

The only remaining step is to impose the closed string conditions to the coordinates and the Lagrange multipliers. Again, $z_1$ is periodic if we impose $\kappa$ to be an integer. Through equation~(\ref{secondrepofz}), we can see that $z$ is automatically periodic because it is a rational function of $z_1$ as long as $n$ is integer. In addition, notice that there is never a branch issue with the square root in the denominator since it is always equal or larger than $1$. Regarding the Lagrange multipliers, their periodicity in the $\sigma$ coordinate is assured provided we enforce periodicity on the $\phi$ and $z$ coordinates.

We want again to finish our description of this solution by studying the conserved charges associated to it for the choice $\phi=w\tau$. This solution has the following non-trivial conserved quantities
\begin{align}
\label{GKP_dispersion}
\notag
	J &= -\pi w T \ , \\
	S &= -\frac{n \kappa T}{4} (n^2-2n -3) \left[ (C^1)^2 + (C^2)^2 \right]  \ , \\
	\notag
	E &= -\frac{\pi w^2 T}{\kappa} + \kappa \left[ \alpha (C^1)^4 - \beta (C^1)^2 (C^2)^2  + \gamma (C^2)^4 \right]  \ ,
\end{align}
where $\alpha$, $\beta$ and $\gamma$ are real functions of $n$ that vanish for $n=2$. Sadly, we were not able to find a closed expression for those functions, but in all the cases we checked they seem to have the same sign.

\subsection{Twisted NR string sector in orbifold compactification}
\label{sec:twisted_states}

As we have discussed in section \ref{sec:Z2orbifold}, the non-relativistic  Nambu-Goto action is invariant under the $\mathbb{Z}_2$ symmetry $z_1 \rightarrow - z_1$.  The transformation $z_1 \rightarrow - z_1$ is not a symmetry  of the non-relativistic Polyakov action, but the transformation 
\begin{eqnarray}
\label{Z2_symm_Polyakov}
z_1 \rightarrow - z_1 \ , \quad
e_{\alpha} \rightarrow (-1)^{s}\, \eb_{\alpha} \ , \quad
\la \rightarrow (-1)^{s}\, \lb \ , \quad
\eb_{\alpha} \rightarrow (-1)^{s} e_{\alpha} \ ,  \quad
\lb \rightarrow (-1)^{s} \la \ , 
\end{eqnarray}
where $s$ is either $0$ or $1$, is a symmetry.  This is a more involved symmetry, since it needs to act simultaneously on the Lagrange multipliers and on the world-sheet Zweibeine. 

The $\mathbb{Z}_2$ symmetry (\ref{Z2_symm_Polyakov}) allows us to identify fields in pairs, namely 
\begin{equation}
\label{fields_identification}
z_1 \sim - z_1 \ ,  \qquad\qquad
\la \sim (-1)^{s}\, \lb \ .
\end{equation}
When making such identification, one needs to keep in mind that also the world-sheet Zweibeine have been identified in pairs, although this is invisible from the world-sheet metric perspective, since the latter is manifestly invariant.  In particular, this implies that the field identification (\ref{fields_identification}) is compatible with imposing conformal gauge.  

The main implication of the field identification (\ref{fields_identification}) is that it allows us to impose boundary conditions on the closed string which are different from the usual ones, and for this reason are called \emph{twisted}.  Therefore, in this section we shall impose on the fields the following boundary conditions
\begin{eqnarray}
\label{twisted_bc}
\notag
z_1(\tau,\sigma) &=& - z_1(\tau, \sigma + 2\pi) \ , \\
\la(\tau, \sigma) &=& (-1)^{s}\,  \lb(\tau, \sigma + 2\pi) \ , \\
\notag
\lb(\tau, \sigma) &=& (-1)^{s} \la(\tau, \sigma+ 2\pi) \ ,
\end{eqnarray}
with the additional condition of swapping the world-sheet Zweibeine 
\begin{equation}
\label{twisted_bc_Zweibeine}
e_{\alpha} (\tau, \sigma)  = (-1)^{s}\, \eb_{\alpha}(\tau, \sigma + 2\pi) \ ,  \qquad\qquad
\eb_{\alpha} (\tau, \sigma)  = (-1)^{s}\,  e_{\alpha}(\tau, \sigma + 2\pi) \ , 
\end{equation}
while for the remaining target space fields $X^{\mu}$ we will either impose the usual closed string boundary conditions (\ref{closed_bc}),  or a twisted boundary condition by using the many additional $\mathbb{Z}_2$ symmetries acting on them.  We remark that these $\mathbb{Z}_2$ symmetries are different from the one acting on $z_1$.  More specifically,  the non-relativistic action is invariant under the additional $(\mathbb{Z}_2)^8$ symmetry which acts independently on the 8 fields as $z_m \rightarrow - z_m,  \phi \rightarrow - \phi, y_i \rightarrow - y_i$. This means we can impose twisted boundary conditions also on them. In particular, for our purpose we shall consider the following boundary condition
\begin{equation}
\label{twisted_bc_zm}
z_m(\tau,\sigma) = - z_m(\tau, \sigma + 2\pi) \ .
\end{equation}
The type of twisted boundary conditions discussed above are sufficient to eliminate the surface term (\ref{surface_term}) which appears in the variation of the action. 

The dispersion relations for the twisted solutions presented below are given again in terms of formulas (\ref{BMN_dispersion}), (\ref{genBMN_dispersion}), (\ref{GKP_dispersion}), as they do not depend directly on the specific boundary conditions imposed and also they have been computed for generic values of integration constants appearing in the solutions. 
In this twisted sector we do not fix $R =1$ as we did before in the closed sector, but we leave it generic and reinstate everywhere.  

\subsubsection{Twisted BMN-like string}
\label{twisted_BMN_sol}

Here we shall consider the BMN-like solution which satisfies the equations of motion and the Virasoro constraints as found in section (\ref{sec:BMN_like_string}), but impose twisted boundary conditions on it.  

If we consider (for simplicity) $\phi = w \tau$, set the constant $\sigma_0 = \pi$, and leave $\kappa$ to be a generic unconstrained real number, then
\begin{eqnarray}
\label{BMN_like_twisted}
\notag
t = \kappa \tau, \qquad\qquad
z_1 = 2R \tan \left[ -\frac{\kappa}{2 R} (\sigma-\pi) \right] \, \qquad
z_m = y_i = 0 \, \qquad
\phi = w \tau \ , \\
\lambda_1 + \lambda_0 = 2\lambda= \frac{w^2}{\kappa} \cos \left[ \frac{\kappa}{R} (\sigma-\pi) \right] \ , \qquad
	\lambda_1 - \lambda_0 = 2\lb = -\frac{w^2}{\kappa} \cos \left[  \frac{\kappa}{R} (\sigma-\pi) \right] \ , 
\end{eqnarray}
is a solution to the equations of motion and to the Virasoro constraints, and it satisfies the twisted boundary conditions (\ref{twisted_bc}) with $s=1$ for $\sigma = 0$, and the closed string boundary conditions for the remaining fields. 

\begin{figure}[h]
\begin{center}
 \includegraphics[scale=1]{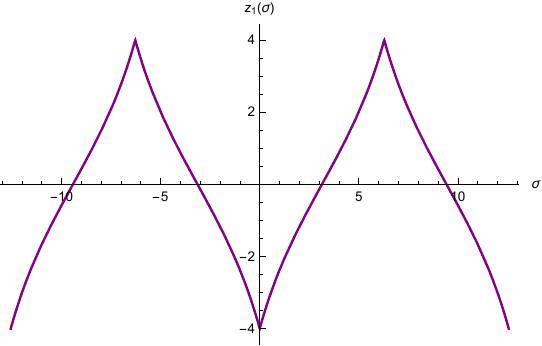}
\end{center}
\caption{Twisted BMN-like solution for values of the parameters $\kappa = 1$ and $R=2$.  The plot shows the functional behaviour of $z_1(\sigma)$ accordingly to the solution (\ref{z1_glued}) obtained by extending  (\ref{BMN_like_twisted}) outside the fundamental domain. } 
\label{z1gluing}
\end{figure}

A possible way\footnote{We thank Alessandro Torrielli for proposing the argument.} to extend the solution (\ref{BMN_like_twisted}) when $\sigma$ spans the whole real line,  so (\ref{twisted_bc}) are satisfied for generic values of $\sigma$,  is by gluing many copies of the solution (\ref{BMN_like_twisted}) outside the fundamental domain $[0, 2\pi]$, namely
\begin{equation}
\label{z1_glued}
z_1= (-1)^n \, 2R \tan \left[ -\frac{\kappa}{2 R} \big(\sigma-(2n + 1)\pi \big) \right]    \qquad\qquad \mbox{if } \sigma \in [2\pi n,  2\pi (n+1)] \ , 
\end{equation}
and the same for the $\lambda, \lb$ solutions
\begin{eqnarray}
\lambda=& \frac{w^2}{2\kappa} \cos \left[ \frac{\kappa}{R} \big(\sigma-(2n + 1)\pi \big) \right]  \qquad\qquad &\mbox{if } \sigma \in [2\pi n,  2\pi (n+1)] \ ,  \\
\lb= & -\frac{w^2}{2\kappa} \cos \left[ \frac{\kappa}{R} \big(\sigma-(2n + 1)\pi \big) \right]  \qquad\qquad &\mbox{if } \sigma \in [2\pi n,  2\pi (n+1)] \ ,
\end{eqnarray}
In the process of gluing, we use the fact that $z_1 \sim -z_1, \lambda \sim - \lb$, and therefore we can make the solution to be continuous at the gluing points, as shown in fig. \ref{z1gluing}.

\begin{figure}[h]
	\begin{center}
		\includegraphics[scale=0.30]{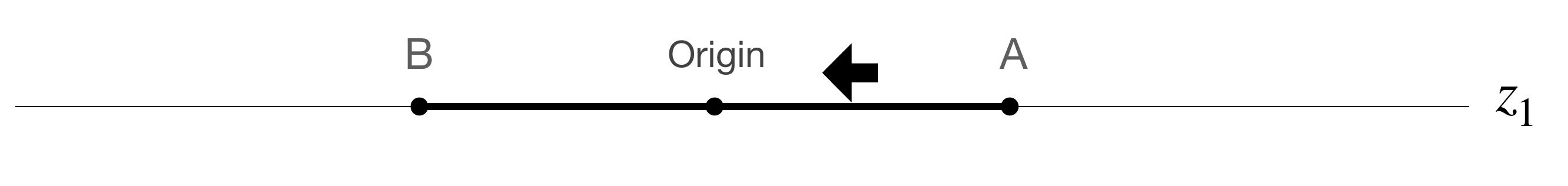}
	\end{center}
	\caption{Twisted BMN-like solution in $z_1$ direction.  The solution represents a segment with end points A and B symmetrically displaced around the origin.  At fixed value of $\tau$,  the motion is parametrically described by $\sigma = 0\rightarrow 2\pi$ and it goes from A to B. } 
	\label{z1_segment}
\end{figure}     

In target space, this solution describes a segment in the $z_1$ coordinate which has initial and final points symmetric with respect to the origin.  At fixed values of $\tau$,  one moves parametrically in $\sigma \in [0,2\pi]$ from the initial point A of the segment to the end point B as described in fig. \ref{z1_segment}.   Since the two points are antipodal, they are identified in the orbifold compactification, and therefore describe a closed twisted string.  The end points A and B have an interpretation of ``reflecting'' walls,  because when $\sigma$ is allowed to span the whole real line and one hits an end point, then one inverts the direction of motion from velocity $v \rightarrow - v$. 
This is responsible for the non-differentiability of the solution in the end points $\sigma_{\text{end}} = 2\pi n$ when $\sigma$ spans the whole line. However, when $\sigma \in [0,2\pi]$ the solution is differentiable in its fundamental domain, and therefore the action well defined.

Because of the twisted boundary conditions,  the ratio $\kappa/R$ does not need to be an integer, as it happens for the untwisted BMN-like solution.  This means that when taking the large $R$ limit,  the $z_1$ profile linearises accordingly to (\ref{largeR_linear}). 
The point $z_1 = 0$, together with $\la = -\lb$, is the \emph{fixed point} of the orbifold.  The solution (\ref{BMN_like_twisted}) is constantly sitting on the point $\la = -\lb$, but not in $z_1 = 0$. Every time it crosses the $z_1 = 0$ point the string gains ``winding'', and the value of $\kappa/R$ contains the information about the winding number.

\subsubsection{Twisted generalized BMN-like string}


We shall consider a generalisation of the twisted BMN-like solution presented in the previous section.  The solution to the equations of motion and Virasoro constraints follows as for the closed sector,  but here only the boundary conditions change.  The solution is 
\begin{eqnarray}
\label{gen_BMN_twisted}
\notag
t = \kappa \tau, \qquad\qquad
z_1 = 2R \tan \left[ -\frac{\kappa}{2 R} (\sigma-\pi) \right] \, \qquad
y_i = 0 \, \qquad
\phi = w \tau \ , 
\end{eqnarray}
where $\kappa$ is an unconstrained real number, and $z_1$ satisfies twisted boundary conditions and extended outside the fundamental domain $\sigma \in [0, 2\pi]$ via the gluing procedure as in formula (\ref{z1_glued}). The same argument also applies to the $\lambda, \lb$ solutions presented below, which must satisfy twisted boundary conditions and are extended outside the fundamental domain via gluing procedure. 
For the $z_m$ we may consider two cases (here $x$ is the dimensionless variable $x= z_1/(2R)$ when $R$ is reinserted):

\vspace{5mm}
\noindent{\bf Periodic boundary conditions on $z_m$}. If we are interested in $z_m$ to satisfy periodic boundary conditions, we need it to depend evenly on $z_1$. This is done by setting $C^1_m = 0$ in (\ref{zmsolution_x_dep}), and it gives
\begin{eqnarray}
z_m &=& R\, C^2_m  \left[ 1 + 4 x \arctan \left(x\right) - x^2 \right] \ , \\
\notag
\lambda_1+\lambda_0&=& 2 \lambda = (\kappa C^2_m C^2_m - \frac{w^2}{\kappa}) \cos \left[ \frac{\kappa}{R} (\sigma-\pi)\right]  \\
&-&\frac{\kappa}{4}C^2_m C^2_m \left[4+4\frac{\kappa^2}{R^2} (\sigma-\pi)^2+\sec^2 \left(\frac{\kappa}{2R^2} (\sigma-\pi)\right)\right] \ ,\\
\notag
\lambda_1-\lambda_0 &=& 2 \lb = (\frac{w^2}{\kappa} - \kappa C^2_m C^2_m) \cos \left[ \frac{\kappa}{R} (\sigma-\pi)\right]   \\
&+&\frac{\kappa}{4}C^2_m C^2_m \left[4+4\frac{\kappa^2}{R^2} (\sigma-\pi)^2+\sec^2 \left(\frac{\kappa}{2R^2} (\sigma-\pi)\right)\right] \ ,
\end{eqnarray}
Notice that the solutions for $\lambda, \lb$ can be extended via the same gluing procedure we used to extend $z_1$ outside the fundamental domain. In addition, we should remark that the gluing procedure keeps the argument of the tangent in $z_1$ inside the interval $[-\frac{\pi \kappa}{2R} , \frac{\pi \kappa}{2R}]$. Extended in this fashion, $\lambda, \lb$ satisfy twisted boundary conditions (\ref{twisted_bc}) with $s=1$, while the $z_m$ solution satisfies periodic boundary conditions with our previous choice of branch for the arctangent, $\arctan (\tan (x))=x$.


\vspace{5mm}
\noindent{\bf Anti-periodic boundary conditions on $z_m$}.  In this case $z_m$ needs to depend oddly in terms of $z_1$, therefore we need to set $C^2_m = 0$, and leads to
\begin{eqnarray}
z_m &=&  R\, C^1_m x \ , \\
\lambda_1+\lambda_0&=& 2 \lambda =  - \frac{w^2}{\kappa} \cos \left[ \frac{\kappa}{R} (\sigma-\pi)\right] - \frac{\kappa}{4}C^1_m C^1_m\ ,\\
\lambda_1-\lambda_0 &=& 2 \lb = \frac{w^2}{\kappa}  \cos\left[ \frac{\kappa}{R} (\sigma-\pi)\right] + \frac{\kappa}{4}C^1_m C^1_m  \ ,
\end{eqnarray}
The solutions above for $\lambda, \lb$, if extended via the gluing procedure,  satisfy twisted boundary conditions (\ref{twisted_bc}) with $s=1$, while the $z_m$ solution satisfies anti-periodic boundary conditions (\ref{twisted_bc_zm}).

\subsubsection{Twisted GKP-like string}

The GKP-like string solutions presented before in the closed sector admits also twisted boundary conditions. The profile for $z_m$ is the same as in (\ref{zm_profile_GKP}), and we consider in particular the case $\theta = 0$ and $\varphi= \omega \tau$, with $\omega = \frac{n\kappa}{2}$ and $n$ integer. We consider the case $\phi = w \tau$ and we need to set the constant $\sigma_0 = \pi$. The solution is 
\begin{eqnarray}
\label{gen_BMN_twisted}
\notag
t = \kappa \tau, \qquad\qquad
z_1 = 2R \tan \left[ -\frac{\kappa}{2 R} (\sigma-\pi) \right] \, \qquad
y_i = 0 \, \qquad
\phi = w \tau \ , 
\end{eqnarray}
where $\kappa$ is an unconstrained real number, $z_1$ satisfies twisted boundary conditions and is extended outside the fundamental $\sigma$ domain with the gluing procedure previously described. For the $\lambda,\lb$ fields we need to impose twisted boundary conditions and their solutions are extended outside the $\sigma$ fundamental domain via the gluing procedure as well. 

Regarding boundary conditions for $z(\sigma)$, we have two possibilities. First we note that its solution (\ref{secondrepofz}) admits a decomposition into even and odd powers of $z_1$. By using Newton's binomial formula one can check that $\left[ (1+ix)^n +(1-ix)^n \right]$ contains only even powers of $x$, while $\left[ (1+ix)^n - (1-ix)^n \right]$ only odd powers of $x$ (where $x$ is the dimensionless variable $x= z_1/(2R)$). 
Therefore we can impose:

\vspace{5mm}
\noindent{\bf Periodic boundary conditions on $z$}. This requires that $z$ depends only on even powers of $z_1$, and therefore we need to set $C^1 = 0$. The solution is: 
\begin{eqnarray}
	z&=&\frac{R\, C^2}{2 \sqrt{\left(1+x^2\right)^n}} \bigg\{ n \left(x^2-1\right)  \left[ \left(1+ix\right)^n +\left(1-ix\right)^n \right]    \notag \\
&+&i 4 x \left[ \left(1+ix\right)^n -\left(1-ix\right)^n \right]   \bigg\} \ , \\
\notag
\kappa (\lambda_1 +\lambda_0) &=&  \kappa^2 z^2 (1+\sec [\frac{\kappa}{R} (\sigma-\pi)]) \frac{4-n^2 (1+\cos [ \frac{\kappa}{R} (\sigma-\pi)]) }{16 R^2}\\
&-&z^{\prime 2} \sec [ \frac{\kappa}{R} (\sigma-\pi)] \cos \, \frac{\kappa (\sigma-\pi)}{4R} - w^2 \cos [\frac{\kappa}{R} (\sigma-\pi)]\ , \\
\notag
\kappa (\lambda_1 -\lambda_0) &=&  -\kappa^2 z^2 (1+\sec [ \frac{\kappa}{R} (\sigma-\pi)]) \frac{4-n^2 (1+\cos [ \frac{\kappa}{R} (\sigma-\pi)]) }{16 R^2} \\
&+& z^{\prime 2} \sec [\frac{\kappa}{R} (\sigma-\pi) ] \cos \frac{\kappa (\sigma-\pi)}{4R} + w^2 \cos [\frac{\kappa}{R} (\sigma-\pi)] \ .
\end{eqnarray}
The solutions for $\lambda, \lb$ above, if extended via the gluing procedure, satisfy the twisted boundary conditions (\ref{twisted_bc}) with $s=1$, while $z_m$ satisfies periodic boundary conditions. 

\vspace{5mm}
\noindent{\bf Anti-periodic boundary conditions on $z$}. In this case $z$ needs to depend only on odd powers of $z_1$, and therefore we need to set $C^2 = 0$. The solution is then:

\begin{eqnarray}
z&=&\frac{R \, C^1}{2 \sqrt{\left(1+x^2\right)^n}} \bigg\{ 4 x \left[ \left(1+ix\right)^n +\left(1-ix\right)^n \right]   \notag \\
&-&i\, n \left(x^2-1\right) \left[ \left(1+ix\right)^n -\left(1-ix\right)^n \right]  \bigg\} \ , \\
\notag
\kappa (\lambda_1 +\lambda_0) &=&  \kappa^2 z^2 (1+\sec [\frac{\kappa}{R} (\sigma-\pi)]) \frac{4-n^2 (1+\cos [ \frac{\kappa}{R} (\sigma-\pi)]) }{16 R^2}\\
&-&z^{\prime 2} \sec [ \frac{\kappa}{R} (\sigma-\pi)] \cos \, \frac{\kappa (\sigma-\pi)}{4R} - w^2 \cos [\frac{\kappa}{R} (\sigma-\pi)]\ , \\
\notag
\kappa (\lambda_1 -\lambda_0) &=&  -\kappa^2 z^2 (1+\sec [ \frac{\kappa}{R} (\sigma-\pi)]) \frac{4-n^2 (1+\cos [ \frac{\kappa}{R} (\sigma-\pi)]) }{16 R^2} \\
&+& z^{\prime 2} \sec [\frac{\kappa}{R} (\sigma-\pi) ] \cos \frac{\kappa (\sigma-\pi)}{4R} + w^2 \cos [\frac{\kappa}{R} (\sigma-\pi)] \ .
\end{eqnarray}

The solutions for $\lambda, \lb$ above, if extended via the gluing procedure, satisfy the twisted boundary conditions (\ref{twisted_bc}) with $s=1$, while $z_m$ satisfies anti-periodic boundary conditions.

\section{Classical NR string solutions in light-cone gauge}
\setcounter{equation}{0}

In this section we shall discuss the twisted BMN-like string solution (\ref{BMN_like_twisted}) as a solution to the string equations of motion derived from the non-relativistic action in the light-cone gauge.  
The reason why we focus on the twisted BMN-like solution is because it is the classical solution around which the semiclassical expansion of the non-relativistic action in light-cone gauge was performed in \cite{Fontanella:2021hcb}.  
The uniform light-cone gauge fixing procedure for the non-relativistic action has been given in \cite{Fontanella:2021hcb}, and we shall refer to that paper for the detail of the derivation.\footnote{Light-cone gauge for non-relativistic actions has also been discussed in \cite{Kluson:2017ufb}.}  However,  here we summarise the key points of the construction.  

Consider the non-relativistic action in the first-order formalism 
\begin{equation}
\label{first_NR_action}
S^{NR} = \int \dd \tau \int_0^{2 \pi} \dd \sigma \, \bigg( p_{\mu} \dot{X}^{\mu} + \frac{\gamma^{01}}{\gamma^{00}} V_1 + \frac{1}{2 T \gamma^{00}} V_2 \bigg) \ , 
\end{equation}
where $p_{\mu}$ is the conjugate momenta, $V_1$ and $V_2$ the linear and quadratic Virasoro constraints respectively. 

Then we introduce light-cone coordinates
\begin{eqnarray}
\notag
X_+ &=& (1-a) t + a \phi \ , \qquad\ \ 
X_- = \phi - t \ , \qquad \\
p_+ &=& (1-a) p_{\phi} - a p_t \ , \qquad
p_- = p_{\phi} + p_t \ , 
\end{eqnarray}
where $a$ parametrises a family of uniform light-cone gauges. We fix uniform light-cone gauge by imposing\footnote{In \cite{Fontanella:2021hcb},  $p_+$ is fixed to a different constant.  However, the choice of this constant does not affect the solution to the equations of motion of our interest, and therefore we can fix it to $1$.}
\begin{equation}
\label{lightcone_gauge}
X_+ = \tau \ , \qquad\qquad p_+ = 1 \ .
\end{equation}
The conjugate momenta to $\lambda_A$ is identically zero,  i.e. $p_{\lambda_A} \approx 0$, and the preservation of this condition in the time evolution provides the further constraint 
\begin{equation}
\partial_{\tau} p_{\lambda_A} = \{ p_{\lambda_A}, \mathcal{H}\} \approx 0 \ ,
\end{equation}
which can be solved by eliminating the Lagrange multipliers.  After solving the two Virasoro constraints $V_1 \approx 0$ and $V_2 \approx 0$ one arrives to an action of the form
\begin{equation}
\label{S_H}
S^{NR} = \int \dd^2 \sigma \, (p_I \dot{X}^I - \mathcal{H} ) \ , 
\end{equation}
where $\mathcal{H} = - p_- (X^I, X'^{I}, p_I)$, i.e. $p_-$ is expressed as a solution in terms of the other variables after solving the constraints.  The index $I$ stands for the light-cone transverse directions. 

At this stage, one can eliminate the transverse conjugate momenta $p_I$ by imposing their Euler-Lagrange equations of motion.
By solving these equations in terms of $p_I$ and substituting them back into (\ref{S_H}) one obtains a light-cone gauge fixed action which solely depends in terms of the light-cone transverse coordinates $X_I$
\begin{equation}
S^{NR} = \int \dd^2 \sigma \, \mathcal{L}_{\text{l.c. gauge}}(X^I, \dot{X}^{I},X'^{I}) \ .
\end{equation} 

If we fix $a=0$, and we make an ansatz of the type
\begin{equation}
z_1 = z_1(\sigma) \ , \qquad\qquad z_m = 0 \ , \qquad\qquad
y_1 = w \tau \ , \qquad
y_2 = y_3 = y_4 = 0 \ , 
\end{equation}
then the only equation of motion that needs to be solved is the one for $z_1$ which reads
\begin{equation}
\frac{(4 R^2 - z_1^2)}{z_1'} \bigg((4R^2 + z_1^2) z_1'' - 2z_1 z_1'^2\bigg)= 0 \ . 
\end{equation}
which admits the only non-trivial solution 
\begin{equation}
z_1(\sigma) = 2 R \tan \bigg( \frac{k}{2R} (\sigma + \sigma_0) \bigg) \ , 
\end{equation}
where $k$ and $\sigma_0$ are constants.  This recovers the (twisted) BMN-like solutions discussed in the conformal gauge. 

On the other hand, if we start with an ansatz of the type 
\begin{equation}
z_1 = 2 R \tan \bigg( \frac{\kappa}{2R} \sigma + c \bigg) \ , \qquad z_m = 0 \ , \qquad
y_1 = f(\tau, \sigma) \ , \qquad
y_2 = y_3 = y_4 = 0 \ , 
\end{equation}
then the only equations of motion that need to be satisfied are the ones for $z_1$ and $y_1$, both of which imply 
\begin{equation}
\ddot{f} - \kappa^2 f'' = 0 \ , \qquad
\Longrightarrow\qquad
f(\tau, \sigma) = f_+(\tau + |\kappa| \sigma) + f_-(\tau - |\kappa| \sigma) \ . 
\end{equation} 
We recover once again that $f$ has to satisfy the wave equation, as we found in the conformal gauge, but with the difference that $\kappa$ plays the r\^ole of the wave propagation speed.  

We remark that it is \emph{not} possible to make the wave propagation speed to be equal to 1 just by a different choice of light-cone gauge, e.g. $X_+ = \ell \tau$, for a certain coefficient $\ell$. The fact that the wave propagation speed must be equal to $\kappa$ is a particular feature of solving the equations of motion in light-cone gauge. For this reason we expect that this solution is the same as the one found in conformal gauge, with the only difference in the wave propagation speed as a hallmark of the gauge choice.  

\section{Conclusions}
\setcounter{equation}{0}

In this paper we found some classical closed string solutions of the NR AdS$_5\times$S$^5$ string theory, which are the NR analogue of the famous BMN and GKP solutions of the relativistic theory. We discussed in detail the fact that the NR string action admits a $\mathbb{Z}_2$ orbifold symmetry, which allows us to impose twisted boundary conditions on solutions. The twisted BMN-like solution is particularly important, because it turns out to be the classical solution around which the semiclassical expansion discussed in \cite{Fontanella:2021hcb} takes place. The expansion in that paper is performed after imposing light-cone gauge, and therefore we checked that the twisted BMN-like solution also satisfies the equations of motion in light-cone gauge, since our whole analysis is done in conformal gauge.  

It is interesting to observe that the equations of motion for the Lagrange multipliers $\lambda_A$ (i.e. $\mathcal{F}^A = 0$) are simple enough to constraint the behaviour of the NR longitudinal coordinates $t$ and $z_1$. As a consequence, it is not possible to have a non-trivial $t$ and a trivial $z_1$ solution. If one of them is non-trivial, then the other one must also be. This is a peculiar feature of the NR theory and it is the reason why the NR BMN-like solution is not as simple as the relativistic one. 

The fact $z_1$ must have a non-trivial profile also makes the semiclassical expansion for  large values of $T$ complicated. This was performed in \cite{Fontanella:2021hcb} with the further assumption of large $R$ in order to shift any $\sigma$-dependent term coming from the classical profile of $z_1$ into higher order corrections.

Although a systematic approach to classical integrability of the NR string action is in progress \cite{Andrea_Stijn}, there are some hints from this work that there might be an underlying integrable structure. This happens for instance in the structure of the equations of motion for the $z_m$ fields in the generalised BMN-like solution. These fields are a solution of a generalized Lamé differential equation, which has the property that the zeroes of their polynomial solutions are described by the solutions to the Bethe equation of the Gaudin Model.

\vskip 1cm
\noindent{\bf Acknowledgements} \vskip 0.1cm
\noindent 

We are in debt of gratitude with A. Torrielli for his collaboration in a related topic and for crucial insights about twisted boundary conditions. 
We are grateful to R. Ruiz and A. Torrielli for reading the manuscript and for providing useful comments and we thank J. Kluso\v{n} for a useful comment on the first version of this paper. AF has been supported by the Deutsche Forschungsgemeinschaft DFG via the Emmy Noether program ``Exact Results in Extended Holography''. JMNG is supported by the EPSRC-SFI grant EP/S020888/1 \emph{Solving Spins and Strings}.
AF thanks Lia for her permanent support. 


\setcounter{section}{0}
\setcounter{subsection}{0}
\setcounter{equation}{0}

\begin{appendices}

\section{Kink solutions and the $\lambda_A$ constant gauge fixing}

In this appendix we will consider the consequences of fixing the residual Diff$_+ \oplus$Diff$_-$ gauge symmetry differently. Instead of fixing $t=\kappa \tau$, we will consider the Lagrange multipliers $\lambda_0$ and $\lambda_1$ to be constant. Although we may think that the solutions we obtain with this choice are gauge-equivalent to the ones we have already obtained, there are some solutions that cannot be transformed into solutions of the conformal gauge with $t=\kappa \tau$.

We need to mention that the solutions presented here are still obscure in some parts. In particular, they require the time coordinate $t$ to be compact, and some of them may not set to zero the surface term (\ref{surface_term}). We comment more on this at the end of the section. However, we still discuss them because of their distinctive behaviour, which reminds solutions of integrable systems.

Let us consider the ansatz:
\begin{align*}
	t&=t(\tau, \sigma) \ , & z_1 &=z_1(\tau, \sigma) \ , & z_m &=0 \ , & y_i &=0 \ , & \phi &=\phi_+ (\tau + \sigma) +\phi_- (\tau - \sigma) \ ,
\end{align*}
then, the equations of motion~(\ref{EoM1}) and (\ref{EoM2}) and the constraints $\mathcal{F}^0$ and $\mathcal{F}^1$ take the form
\begin{align}
		&\frac{z_1}{1- (\frac{z_1}{2})^4} \left( \lambda_1 z'_1 -\lambda_0 \dot{z}_1 \right)=0 \ , &	&t' + \frac{z'_1}{1 + (\frac{z_1}{2R})^2}=0 \ , \\
	&\frac{z_1}{1- (\frac{z_1}{2})^2} \left( \lambda_1 t' -\lambda_0 \dot{t} \right)=0 \ , &	&\dot{t} + \frac{\dot{z}_1}{1 + (\frac{z_1}{2R})^2}=0 \ .
\end{align}
As we already know, we can solve the two rightmost equations by choosing
\begin{equation}
	t=-2 f(\tau + \sigma) + 2 g(\tau - \sigma) \ ,  \qquad \qquad z_1= 2 \tan [f(\tau + \sigma) + g(\tau - \sigma)+c] \ .
\end{equation}
If we substitute this solution into the two leftmost equations, we find three possible choices: either we choose $\lambda_0=\lambda_1=0$, $\lambda_0=\lambda_1$ with $g=0$, or $\lambda_0=-\lambda_1$ with $f=0$. The first solution is equivalent to the BMN-like string we have studied with $\phi=0$, but the other two cannot be obtained from our previous gauge choice. This is because both $f$ and $g$ transform as scalars under the residual Diff$_+ \oplus$Diff$_-$ world-sheet diffeomorphisms, meaning that we cannot change a non-constant function into a constant one.

\begin{figure}[t]
\begin{center}
 \includegraphics[width=0.45\textwidth,keepaspectratio]{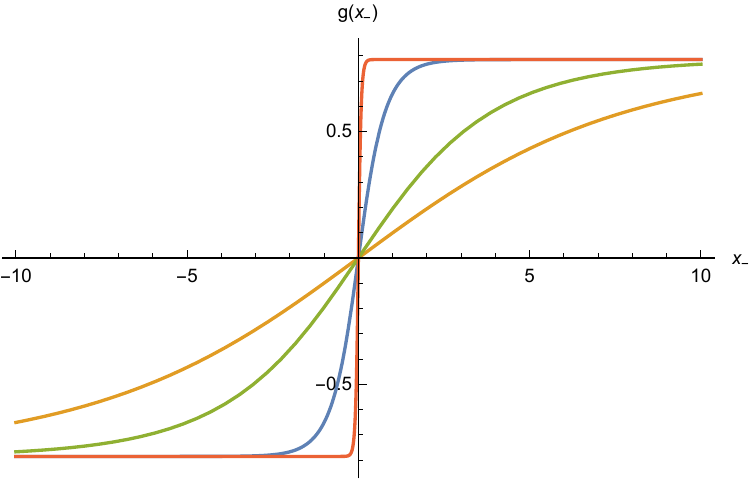}
\end{center}
\caption{Kink solution for $\phi=m x_-$ for different values of $\lambda_1/m^2$.} 
\label{figurekink}
\end{figure}

If we move our attention now to the Virasoro constraints, we find
\begin{align}
	V_1 + V_2 \propto (\lambda_0 + \lambda_1 ) f' -\cos[2 (f+g)] \phi_+^{\prime 2} \ , \\
	V_1 - V_2 \propto (\lambda_0 - \lambda_1 ) g' +\cos[2 (f+g)] \phi_-^{\prime 2} \ ,
\end{align}
where primes denote derivative with respect to the appropriate light-cone variable. If we consider now the case $\lambda_0=\lambda_1=0$, the Virasoro constrains forces us either to set $f+g=\pi/2 + m\pi$ with $m$ integer or to set $\phi=0$. However, if we consider either of the other two cases, we can find solutions with non-trivial $t$ and $\phi$. As one case can be mapped to the other by exchanging the sign of $\lambda_0$ and of the $\sigma$ variable, let us consider $f=\lambda_0+\lambda_1=0$ for concreteness. Then, the solution has to fulfil the equation
\begin{equation}
	2\lambda_0 g'+\cos (2g) \phi_-^{\prime 2}=0 \Longrightarrow g(\tau-\sigma)=\pm \arccos \left( \pm \frac{1+\exp [\frac{1}{\lambda_1} \int_{x^0_-}^{\tau-\sigma} \phi_-^{\prime 2} (x) dx ]}{\sqrt{2+2\exp [\frac{2}{\lambda_1} \int_{x^0_-}^{\tau-\sigma} \phi_-^{\prime 2} (x) dx ]}} \right) \ .
\end{equation}
where $x^0_-$ is an integration constant and the two $\pm$ signs are uncorrelated. In what follows we define $x_{\pm} = \tau \pm \sigma$. Let us analyse some particular cases of interest. We will start first by fixing $\phi_-=m x_-$, which implies
\begin{equation}
	m^2 \cos (2g)-2\lambda_1 g'=0 \Longrightarrow g(x_-)= \arccos \left( \frac{1+e^{\frac{2m^2 (x_-+x_-^0)}{\lambda_1}}}{\sqrt{2+2e^{\frac{4m^2 (x_-+x_-^0)}{\lambda_1}}}} \right) \ ,
\end{equation}
where $x^0_-$ is an integration constant. Here we have to be careful choosing the branch of the arc-cosine in such a way that $g$ and its derivative are continuous. This can be done by choosing it as $2\pi n \pm$ sgn$(x_-+x^0_-) |\arccos(\dots)|$, with $n$ an integer. In addition, as $t$ is proportional to $g$, we will only consider the solution with plus sign in front, as we would like $\dot{t} \geq 0$. The particular form of $g$ is a kink-like function that interpolates between $-\pi/4$ and $\pi/4$, as can be seen in figure~\ref{figurekink}.

\begin{figure}[t]
\begin{center}
 \null  \hfill  \includegraphics[width=0.45\textwidth,keepaspectratio]{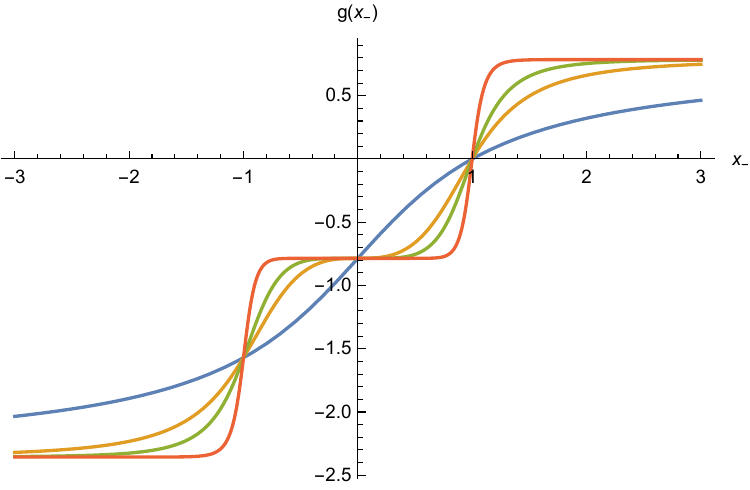} \hfill  \includegraphics[width=0.45\textwidth,keepaspectratio]{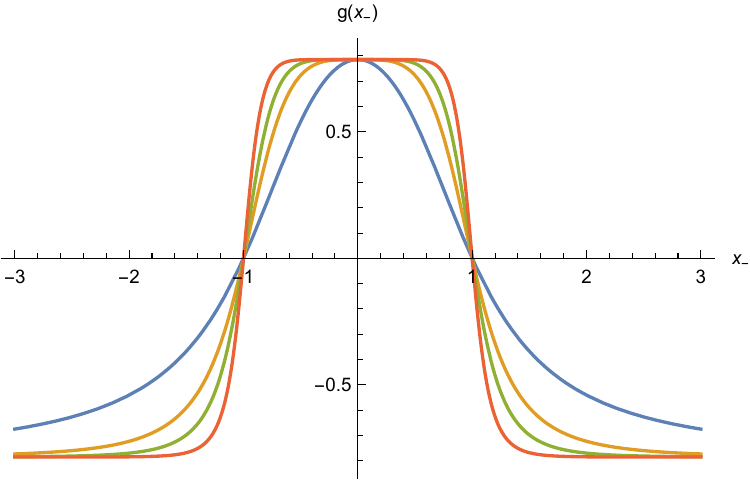}  \hfill \null
\end{center}
\caption{Kink solution for $\phi=\sqrt{m x_-}$ for different values of $\frac{m}{2\lambda_1}$. The left image corresponds to odd values while the right one corresponds to even values.} 
\label{figurekink2}
\end{figure}

A second interesting case would be given by $\phi_-=\sqrt{|m x_-|}$, which fixes
\begin{equation}
	m^2 \cos (2g)-2x_- \lambda_1 g'=0 \Longrightarrow g(x_-)= \arccos \left( \frac{1+A x^{\frac{m}{4\lambda_1}}_-}{\sqrt{2+2A^2 x^{\frac{m}{2\lambda_1}}_-}} \right) \ ,
\end{equation}
where $A$ is an integration constant. This result can be obtained by solving the differential equation for $\phi_-=\sqrt{m x_-}$ and gluing the solution for $m$ for $x_->0$ with a solution for $-m$ for $x_-<0$.\footnote{It seems this is not necessary, as the solution is invariant under the substitution $x_- \rightarrow 1/x_-$.}  The arc-cosine is a little more involved, but it still follows the rule that $g$ should be differentiable. In this case we will find a different behaviour depending on the value of the inverse characteristic distance, $\frac{m}{2\lambda_1}$, which has to be an integer. If this value is odd, we will find a kink between $-3\pi /4 $ and $\pi /4$ (actually, it looks more and more like the composition of two kinks as the characteristic distance decreases). If the value is even, we will find a breather-like behaviour that starts at $-\pi/4$, grows to $\pi/4$, and comes back (Figure \ref{figurekink2}).

A third case of interest would be when both functions are the same
\begin{equation}
	g=\phi_-=\frac{\arcsin [4 \lambda_1 (x_- - x^0_-) ]}{2} \ .
\end{equation}
where $x_-^0$ is an integration constant. Notice that the functions are only real when $|4 \lambda_1 (x_- - x_-^0)| \leq 1$. Thus, the solution is not defined for the whole range of $\tau$ and $\sigma$.

The last case of interest we will study is obtained by setting $g=m x_-$, which gives us
\begin{equation}
	\phi_-=A\pm\sqrt{\frac{2 \lambda_1}{m}} \, F(m (x_- - x_-^0) ,2) \ ,
\end{equation}
where $A$ and $x_-^0$ are integration constants and $F$ is the incomplete elliptic integral of first kind. Similarly to the previous case, the solution is real only if $|4 m (x_- - x_-^0)|\leq \pi$, as the complete elliptic integral of first kind becomes imaginary if the elliptic parameter is greater than unity.\footnote{This happens because the incomplete elliptic integral of first kind can be brought to elliptic parameter lower than unity, where it is well defined, as $F(\phi , m) = F(\theta, 1/m)/\sqrt{m}$ with $\sin \phi= \sqrt{m} \sin \theta$ and $0\leq m \leq 1$. Thus, if we explore the regime of $\phi \geq \arcsin \sqrt{m}$, we will find imaginary values of $\theta$.}

It is still unclear whether the solutions presented above set to zero the surface term (\ref{surface_term}). In the assumption that the world-sheet is decompactified, hence $\sigma$ spans the real line, only the breather-like solution seems to set the surface term to zero. The fact that all these solutions are constant at infinite values of $x_-$ allows us to set to zero the first term of (\ref{surface_term}), which enters in a similar way also in relativistic string theory, but not the second term, which contains the Lagrange multipliers and is typical of NR string theory.

\end{appendices}


\end{document}